\documentclass[reprint,amsmath,amssymb,aps,superscriptaddress, nofootinbib,prl]{revtex4-2}

\usepackage[T1]{fontenc}

\usepackage{multibib}           
\newcites{S}{Supplementary References}
\makeatletter                   
\AtBeginDocument{\let@environment{thebibliography}{rtx@thebibliography}}
\makeatother

\usepackage{graphicx} 
\usepackage{braket}
\usepackage{mathtools}
\usepackage{bbm}
\usepackage{geometry}
\usepackage{url}
\usepackage{amsthm}

\usepackage{tabularray}
\usepackage{multirow}
\usepackage{booktabs}       
\usepackage{threeparttable} 
\usepackage{siunitx}        

\usepackage{comment}
\usepackage{tikz}
\usepackage{quantikz}
\usepackage{relsize}

\usepackage{thm-restate}
\newtheorem{theorem}{Theorem}

\newtheorem{defn}{Definition}
\newtheorem{cor}{Corollary}

\usepackage{array}
\usepackage{bm}

\usepackage[normalem]{ulem} 

\usepackage[utf8]{inputenc}
\usepackage[T1]{fontenc}
\newcommand{\paren}[1]{\left(#1\right)}

\usepackage[colorinlistoftodos, textsize=small, color=orange!50]{todonotes}

\newcommand{\norm}[1]{\left\lVert#1\right\rVert}
\newcommand{\ceil}[1]{\left\lceil #1 \right\rceil}
\newcommand{\floor}[1]{\left\lfloor #1 \right\rfloor}

\newcommand{\be}{\textsc{Be}}

\usepackage{xcolor}
\definecolor{pennylaneblue}{RGB}{14,170,249}
\definecolor{pennylanedarkblue}{RGB}{64,80,187}
\definecolor{XanaduBlue}{HTML}{4D53C8}
\usepackage[colorlinks=true, allcolors=pennylaneblue]{hyperref}
\usepackage{orcidlink}
\usepackage{cleveref}

\usepackage[table]{xcolor}

\usepackage{tikz}
\usetikzlibrary{positioning, fit, backgrounds, shapes.geometric, calc, arrows.meta}

\definecolor{prxblue}{RGB}{55, 113, 172}
\definecolor{prxred}{RGB}{204, 51, 17}
\definecolor{prxpurple}{RGB}{119, 51, 153}

\makeatletter
\renewcommand\subsection{\@startsection{subsection}{2}{\z@}%
  {-1.5ex \@plus -0.5ex \@minus -0.2ex}
  {0.1ex}
  {\normalfont\normalsize\bfseries\raggedright}}
\renewcommand\subsubsection{\@startsection{subsubsection}{3}{\z@}%
  {-1.2ex \@plus -0.3ex}%
  {-1em}
  {\normalfont\normalsize\bfseries\itshape}}
\makeatother

\begin{document}

\preprint{APS/123-QED}

\title{Quantum Simulations for Extreme Ultraviolet Photolithography}


\author{Tyler D. Kharazi}
\affiliation{Xanadu, Toronto, ON, M5G2C8, Canada}
\affiliation{Department of Chemistry, University of California, Berkeley,
Berkeley, California 94720, USA}
\author{Stepan Fomichev}
\affiliation{Xanadu, Toronto, ON, M5G2C8, Canada}
\author{Shu Kanno}
\affiliation{Science \& Innovation Center, Mitsubishi Chemical Corporation, 1000, Kamoshida-cho, Aoba-ku, Yokohama 227-8502, Japan}
\author{Takao Kobayashi}
\affiliation{Science \& Innovation Center, Mitsubishi Chemical Corporation, 1000, Kamoshida-cho, Aoba-ku, Yokohama 227-8502, Japan}
\author{Juan Miguel Arrazola}
\affiliation{Xanadu, Toronto, ON, M5G2C8, Canada}
\author{Qi Gao}
\affiliation{Science \& Innovation Center, Mitsubishi Chemical Corporation, 1000, Kamoshida-cho, Aoba-ku, Yokohama 227-8502, Japan}
\author{Torin F. Stetina}
\affiliation{Xanadu, Toronto, ON, M5G2C8, Canada}

\begin{abstract}
A key challenge of extreme ultraviolet (EUV) lithography in semiconductor fabrication is the line edge roughness or ``blur'' produced by the electron cascades following absorption of a high-energy photon. Here we present quantum algorithms to compute EUV absorption and photoelectron emission spectra, which are key to predicting blur. The first is a time-domain algorithm resolving absorption at a given frequency; the second is a first-quantized plane-wave algorithm computing the photoemission spectrum via real-time dynamics that treats bound and continuum states on equal footing. For a model photoresist monomer IMePh, 92 eV absorption requires $200$ logical qubits and $10^{9}$ non-Clifford gates per circuit with $10^3$ shots, while the photoemission spectrum needs $\geq 10^{14}$ gates, $10^4$ shots, and several thousand logical qubits. These results establish high-fidelity quantum simulations as a key component to parameterize the multi-scale macroscopic models required to overcome the electron blur bottleneck in semiconductor miniaturization.
\end{abstract}

\maketitle


\section*{Introduction}
Extreme Ultraviolet (EUV) photolithography underpins the manufacturing of next-generation semiconductor devices, yet further reducing the size of transistors is fundamentally constrained by line edge roughness and ``blur'' inherent to the EUV exposure mechanism. Unlike deep ultraviolet (DUV) lithography driven by direct photochemistry, EUV patterning relies on a multi-stage electronic cascade. The absorption of a 92 eV photon triggers an immediate attosecond-scale photoionization, followed by a relaxation phase spanning femtoseconds where Auger decay and inelastic scattering generate a shower of low-energy secondary electrons~\cite{kozawa2010radiation, ogletree2016molecular, ma2020investigating}. It is this secondary population, in addition to the primary photoelectron, that both significantly drive the chemical transformation of the photoresist. Since the spatial extent of this electron cascade competes with the intended nanometer feature sizes, the predictive design of photoresists requires accurate simulation of the high-energy decay channels, specifically photoabsorption and Auger electron emission, that define the initial conditions for this complex process over multiple scales; spatially, and temporally~\cite{kim2018multiscale, lee2021multiscale}.
\begin{figure*}[htp]
    \centering
    \includegraphics[width=1.0\linewidth]{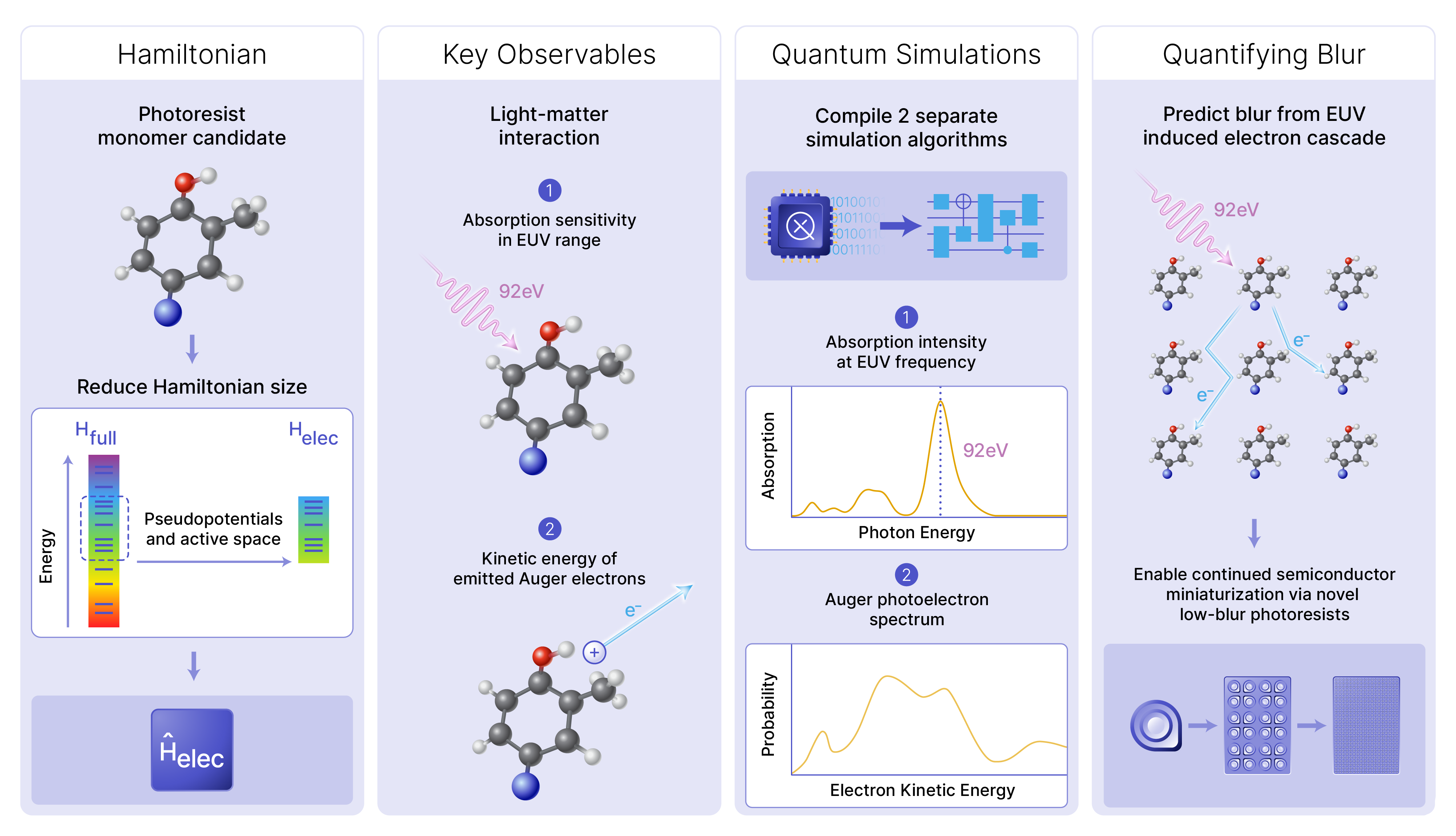}
    \caption{\textbf{Proposed workflow for \emph{ab initio} design of low-blur EUV photoresists.} The framework connects atomic-scale electronic structure to macroscopic lithographic performance through four stages: (Hamiltonian) A candidate monomer, such as 4-iodo-2-methylphenol, is selected. Its full Hamiltonian ($H_\text{full}$) is dimensionally reduced to an effective electronic Hamiltonian ($H_\text{elec}$) utilizing pseudopotentials and/or active space selection to enable efficient simulation in a chosen basis and representation. (Key Observables) Two critical properties governing EUV lithography are targeted: the absorption sensitivity at the specific 92 eV operating frequency and the kinetic energy spectrum of secondary electrons generated via Auger decay. (Quantum Simulations) Two distinct quantum algorithmic primitives are compiled: a coherent time-domain approach to estimate the dipole autocorrelation function for absorption sensitivity, and a real-time electronic dynamics simulation to resolve the kinetic energy distribution of electrons ejected into continuum states. (Quantifying Blur) These quantum observables parameterize the initial conditions of electron cascades in larger bulk materials, allowing for the computational predictions of stochastic ``electron blur'' which ideally enables the design of materials that support continued semiconductor miniaturization.}
    \label{fig:hero}
\end{figure*}

The photoabsorption cross-section at the EUV operating energy (92 eV) serves as the primary metric for screening photoresist candidate materials. Since this absorption event initiates the secondary electron cascade essential for lithographic patterning, the ability to reliably rank molecules by their sensitivity is a critical engineering objective. However, effective screening is currently obstructed by the computational complexity of resolving the many high energy states required to describe 92 eV excitations, as well as the electron correlation present in said excited states, requiring treatment beyond density functional theory (DFT) in some materials~\cite{closser2017importance}. While accurate results are theoretically possible via correlated wavefunction based methods given a sufficiently large diffuse basis beyond DFT, the classical cost of solving for the thousands of correlated interior eigenstates is prohibitive while attempting to retain control over the accuracy of the computed results. 


Next, accurately predicting the low-energy Auger electron kinetic energy spectrum is a prerequisite for parameterizing the multi-scale models used to design EUV photoresists~\cite{ma2020investigating, kostko2018fundamental, fernandez2024molecular, theofanis2020modeling}. A large portion of the lithographic chemistry in these systems is driven primarily by the cascade of low-energy secondary electrons rather than the initial high-energy photoelectron. Consequently, the emitted electron yield and its kinetic energy distribution are key parameters for determining the ``dose-to-clear'' and ultimate resolution of the etched material. However, choosing materials that increase EUV absorption using heavy-atom substitutions, such as iodinated organic photoresist monomers like 4-iodo-2-methylphenol (IMePh), has shown that DFT based simulations are not able to accurately model this process~\cite{ma2020investigating}. Specifically, these methods failed to reproduce the intense low-kinetic-energy spectral features observed experimentally, which are explicitly identified as Auger electrons. 
In general, the key target observable is the kinetic energy spectrum of a photoemitted electron which could correspond to either photoionization or Auger decay processes. Even though the previously mentioned DFT method was not accurate, there are other classical approaches that attempt to incorporate electron correlation and continuum effects in the literature, such as Feshbach-Fano approaches~\cite{skomorowski2021feshbach, matz2022molecular, jayadev2023auger}, R-matrix theory~\cite{burke2007r,descouvemont2010r,brown2020rmt,mavsin2020ukrmol+}, and explicit real-time Green's function based methods~\cite{covito2018real, covito2018benchmarking}. The first two methods rely on approximate treatments of the continuum states, and the explicit time evolution approaches are typically constrained to small model systems or perturbative treatment of electron correlation due to the computational complexity of the full many-body dynamics. 

This simulation gap, which hinders the accurate parameterization of larger multi-scale models, stems from multiple causes. First, standard electronic structure methods typically rely on spatially localized basis sets, typically Gaussian orbital bases, that cannot accurately describe high-energy resonances embedded in the ionization continuum \cite{aaberg1982theory, reinhardt1982complex, jagau2022theory}. While plane-wave bases can more naturally represent the continuum states, classically simulating this in a sufficiently large volume is computationally prohibitive. Second, Auger electrons emitted from the EUV excited system form much slower in time than direct photoexcitation or photoemission, as a real-time simulation approach must evolve the correlated many-electron system over the femtosecond scale to observe Auger decay. 

In this work, we present two quantum simulation algorithms that compute key observables used in models predicting EUV-induced blur in photoresists. A summary of the results and workflows presented in this paper is shown in Fig.~\ref{fig:hero}. The first observable is the EUV absorption sensitivity, or more specifically, the photoabsorption cross-section at 92 eV. For this task, we present a novel optimization of standard quantum algorithms for computing electronic spectra, based on a coherent time-domain representation tailored towards single frequency absorption sensitivity. The second observable is the kinetic energy spectra of electrons ejected from the target monomer, also referred to as the photoelectron emission spectra, or Auger spectra. For simplicity, we will refer to this as the \emph{photoemission spectra} as a shorthand from here on. For this task, we present a first-quantized quantum simulation approach to efficiently compute the Auger spectra while accurately treating bound and continuum effects on equal footing. For both algorithms, we provide quantum resource estimates using the IMePh monomer as a model system, showcasing the kind of quantum computer capabilities that would be needed to execute these algorithms in practice and thereby establishing a concrete use case of quantum computing in the semiconductor industry. 


\section*{Results}\label{sec:theory}


For all results introduced in this paper, we consider the electronic structure model for a system of $\eta$ electrons and $L$ nuclei under the Born-Oppenheimer approximation, where we have clamped nuclei treated as classical point charges, $\zeta_l$, fixed at positions $\mathbf{R}_l \in \mathbb{R}^3$. Due to the short timescales of initial EUV excitation and electronic dynamics, this is an accurate approximation due to the fact that the nuclei are approximately static. Additionally, unless otherwise noted, we assume usage of atomic units throughout this work, $\hbar = 4\pi \epsilon_0=e=m_e=1$ where $\epsilon_0$ is the permittivity of free space, $e$ is the electron charge, and $m_e$ is the electron mass.

The electronic Hamiltonian is given by the contribution of four terms, the kinetic energy $T$, electron-nuclear attraction $U$, and electron-electron repulsion $V$,
with a constant factor shift from the classical nuclear-nuclear repulsion energy, $V_\text{nuc}$,  detailed below. The definition of the Hamiltonian is then
\begin{align}\label{eq:base_coulomb_ham}
    H  = -&\sum_i^{\eta} \frac{\nabla^2_i}{2} + \sum_{l}^{L} \sum_i^{\eta} \frac{-\zeta_l}{|\boldsymbol{R}_l - \boldsymbol{r}_i |} \nonumber \\ &+ \frac{1}{2}\sum_{i \neq j}^{\eta} \frac{1}{|\boldsymbol{r}_i - \boldsymbol{r}_j |} + V_\text{nuc},
\end{align}
where $\nabla^2_i$ is the spatial Laplacian for particle $i$, and $\boldsymbol{r}_i \in \mathbb{R}^3 $ is the position of particle $i$.

While the Hamiltonian in Eq.~\eqref{eq:base_coulomb_ham} is the general \emph{ab initio} model of non-relativistic electronic structure theory agnostic to a given basis, the distinct physics of photoabsorption and Auger decay use different assumptions and restrictions on the choice of an explicit basis. For the initial photoabsorption event, in which we target transitions between localized bound and Rydberg type states, we employ a compact second-quantized formulation that efficiently captures electron correlation within a chemically relevant active space. Conversely, the subsequent Auger decay involves the ejection of electrons into high-energy scattering states, a process ill-suited for standard localized Gaussian basis sets, often found in quantum chemistry calculations. To rigorously model this continuum dynamics, we use a first-quantized plane-wave basis representation that naturally supports delocalized electronic wavefunctions, but typically requires a much larger basis to represent the physics accurately. 

\subsection*{Photoresist Sensitivity}
\subsubsection*{Background} In the first step of the EUV lithographic process, a focused light source at 92 eV is applied to the photoresist material. Provided that the photoresist has a significant absorption cross section at this excitation frequency, some of the incident light will be absorbed, leading to photoexcitation and potentially ionization of a relatively tightly bound, semi-core electrons, similar to X-ray absorption-like excitations. This absorption cross-section directly determines the key performance metric of the photoresist, its so-called \textit{sensitivity}: any material candidate whose absorption cross-section is too low will not be an effective photoresist material for use in the photolithography process. This requirement then directly translates into one of our two key material design tasks: using a quantum simulation algorithm, calculate the EUV absorption cross-section at 92 eV, to determine whether the material candidate passes the EUV sensitivity threshold.  

While EUV light is high-energy, it is still well described within the dipole approximation. This means we can reliably treat the photoexcitation in the semi-classical picture, where the incoming photon is treated as a classical perturbing electric field, while electrons in the photoresist are treated quantum mechanically. 

Much as in the case of X-ray or ultraviolet-visible range absorption spectroscopy, the EUV absorption cross-section may be written using the Kramers-Heisenberg relation in atomic units as
\begin{equation}
    \sigma_A(\omega) = \frac{4\pi}{3 c}\omega \sum_{F \neq I} \sum_{\mu = x, y, z}\frac{\left| \bra{\Psi_F}D_{\mu}\ket{\Psi_I} \right|^2 \gamma}{((E_F - E_I) - \omega)^2 + \gamma^2},
    \label{eq:crosssection}
\end{equation}
where $\ket{\Psi_I}$, $\ket{\Psi_F}$ are the initial and final many-body eigenstates of the photoresist molecule's Hamiltonian $H$ with associated eigenvalues $E_I, E_F$, and $\omega$ is the frequency of incoming radiation, $c$ is the speed of light, $\mu$ runs over the three Cartesian directions of the electric dipole operator $D_{\mu}$, and $\gamma$ is the absorption line broadening. 

For the purpose of quantum simulation, we reformulate this observable in terms of the imaginary part of the following Green's function
\begin{equation}
    \mathcal{G}_\mu(\omega) = \bra{\Psi_I} D_{\mu} \frac{1}{H - E_I 
    - \omega + i\gamma} D_{\mu} \ket{\Psi_I}.
    \label{eq:g-freq-unnormalized}
\end{equation}
By substituting the resolution of identity between the Hamiltonian resolvent $(H - E_I - \omega + i\gamma)^{-1}$ and the dipole operator $D_{\mu}$ and then taking the imaginary part, we can readily recover Eq.~\eqref{eq:crosssection}, up to multiplication by scalars. The photoresist Hamiltonian entering this definition may be represented either in first or second quantization: in this section, we find it convenient to express it in the latter form, namely 
\begin{align}\label{eq:el-ham}
    H_\text{SQ} &= E + \sum_{p,q = 1}^N  (p|\kappa|q) a_{p}^\dagger a_{q} \nonumber \\ &+ \frac{1}{2}\sum_{p,q,r,s=1}^N (pq|rs) a_{p}^\dagger a_{r}^\dagger a_{s}  a_{q},
\end{align}
where $a_{p}^{(\dagger)}$ is the annihilation (creation) operator for spin-orbital $p$, $E$ is the energy offset, $N$ is the number of spin-orbitals, and $(p|\kappa|q)$ and $(pq|rs)$ are the one- and two-electron integrals, respectively, defined in chemist notation as
\begin{align}
    (pq|rs) &\equiv \iint_{\mathbb{R}^6} d\mathbf{r}_1 d\mathbf{r}_2 \phi_p(\mathbf{r}_1) \phi_q(\mathbf{r}_1) \frac{1}{r_{12}} \phi_r(\mathbf{r}_2) \phi_s(\mathbf{r}_2)\\
    (p|\kappa|q) &\equiv \int_{\mathbb{R}^3} d\mathbf{r}_1   \phi_p(\mathbf{r}_1) \left(-\frac{1}{2}\nabla_{\mathbf{r}_1}^2 -\sum_A \frac{\zeta_A}{r_{1A}} \right)\phi_q(\mathbf{r}_1)  ,
\end{align}
with $\phi_p(\mathbf{r})$ being the $p$'th spatial molecular orbital, $\zeta_A$ being the atomic number of the $A$'th atom, and $r_{12} = |\mathbf{r}_1 - \mathbf{r}_2|, r_{1A} = |\mathbf{r}_1 - \mathbf{r}_A|$ the distances between electronic positions $\mathbf{r}_i$ and nuclei positions $\mathbf{R}_A$. \\ 

\subsubsection*{Coherent time-domain algorithm} Unlike in many prior works on spectroscopy, the specifics of the EUV application are such that rather than needing access to the entire spectrum, as in a fingerprinting application \cite{fomichev2024simulating,fomichev2025fast,kunitsa2025quantum}, or computing cumulative absorption over a region \cite{zhou2025quantum}, what is required is an accurate absorption cross-section at a single, well-defined frequency. Given this, here we propose a quantum simulation algorithm that focuses on only getting this specific frequency dependent response, reducing the use of quantum resources relative to reconstructing the entire spectrum with full fidelity. 

The most direct approach would involve constructing the resolvent $(H - E_I - \omega + i \gamma)^{-1}$ and then evaluating the matrix element in Eq.~\eqref{eq:g-freq-unnormalized}. However, constructing matrix inverses is typically an expensive task on a quantum computer. Instead, we leverage the time-domain formulation of the same matrix element that allows us to cast the same observable directly in terms of a (coherent) quantum simulation. 

First, we re-write Eq.~\eqref{eq:g-freq-unnormalized} using the discrete-time Fourier transform
\begin{equation}\label{eq:time-domain}
    \mathcal{G}_\mu(\omega) = \frac{\tau}{2\pi}\sum_{j = -\infty}^{\infty}  e^{i\omega\tau j} \underset{\mathcal{G}_\mu(\tau j)}{\underbrace{e^{-\gamma \tau |j|}\bra{\Psi_I}  D_\mu e^{-i H \tau j}  D_\mu \ket{\Psi_I}}},
\end{equation}
following Refs. \cite{fomichev2024simulating,fomichev2025fast,linHeisenbergLimitedGroundStateEnergy2022}. Once cast in this form, we can further re-write it in a suggestive form
\begin{multline}\label{eq:mathcal_g}
    \mathcal{G}_\mu(\omega) = \\
    \bra{\Psi_I} D_\mu \underbrace{\left[ \sum_{j=-j_\text{max}}^{j_\text{max}} p_{\omega,\gamma,\tau}(j) \left( e^{-i\hat H \tau} \right)^j \right]}_{\equiv  G_t}  D_\mu \ket{\Psi_I},
\end{multline}
where we define $p_{\omega,\gamma,\tau}(j) = (\tau / 2\pi) e^{-\gamma \tau |j| + i\omega \tau j}$ and introduced a cutoff $j_\text{max}$. In this form, it is clear that for a finite range of $j$, the operator in the middle amounts to a polynomial of the one-step time evolution operator, and is in fact naturally a linear combination of unitaries (LCU). Implementing this operator and then performing, for example, a Hadamard test for the specific value of $\omega = \omega_\text{EUV}$ would give direct access to the cross-section of interest in evaluating EUV sensitivity.

There are different strategies one could employ for preparing this polynomial operator: one approach could be to do so directly using the standard \textsc{prepare}-\textsc{select}-\textsc{prepare} construction of an LCU (more details in the Supplementary Information S2-A). However, here we instead opt to use generalized quantum signal processing (GQSP)~\cite{motlagh2024generalized}, where the individual time evolution unitaries will be implemented using a product formula. The high-level approach of the algorithm is as follows
\begin{enumerate}
    \item Classically compute an approximate $D_{\mu} \ket{\Psi_I}$ state and normalization factor $\mathcal{N} = ||D_{\mu} \ket{\Psi_I}||$.
    \item Initialize $\frac{D_{\mu}}{\mathcal{N}} \ket{\Psi_I}$ state on the quantum computer using the sum-of-Slaters (SOS) method~\cite{fomichev2024initial}.
    \item{Use the GQSP  algorithm to compute $G_t$ applied to the initial state with a polynomial approximation of $p_{\omega,\gamma,\tau}(j)$ for a given timestep propagator $e^{-iH\tau}$ and $j_{\text{max}}$ cutoff. The propagator is implemented using a Trotter product formula.}
    \item{Measure $\mathcal{G}_{\mu}(\omega)$ in Eq.~\eqref{eq:mathcal_g} using the Hadamard test for a given $\epsilon$ error in the absorption cross section. }
\end{enumerate}

In the first step of the algorithm, we may choose from any standard technique for preparing an approximate ground state on a classical computer, such as the density matrix renormalization group (DMRG) method~\cite{white1992density, chan2002highly} or selected configuration interaction \cite{bender1969studies,whitten1969configuration}. We assume the use of a Gaussian type orbital (GTO) basis set, but the algorithm will work for any choice of basis. 

For the second step of the algorithm, we take the classical ground state wavefunction, and apply the dipole operator onto it, which can be efficiently performed by virtue of the fact the dipole operator is a one-body operator. We then take this dipole-acted state and prepare it on a quantum computer using the sum-of-Slaters (SOS) method~\cite{fomichev2024initial}. The SOS state preparation approach for second quantized Hamiltonians in a GTO basis typically costs orders of magnitude fewer non-Clifford gates than the time evolution operator in the following step, so it is not a bottleneck for this algorithm~\cite{fomichev2025fast}.

In the third step, we construct and implement the time evolution operator, which is the main source of the number of non-Clifford gates in the full algorithm. A key part of this step is the implementation of the individual time evolution unitaries $e^{-iH\tau}$ for a single time step $\tau$.  Rather than pursuing a qubitization-based strategy which would necessarily incur a dependency on the full one-norm of this Hamiltonian, we instead leverage compressed double factorization (CDF)~\cite{cohn2021quantum, yen2021cartan, oumarou2024accelerating} and a second order Trotter product formula to implement the unitary, following the approach of Ref. \cite{fomichev2025fast}. This is described in more detail in the Methods section. 

Once the framework for implementing the timestep propagator $U(\tau) = e^{-iH \tau}$ is constructed in the way described above, we need to implement $ G_t = \sum_j p_{\omega,\gamma,\tau}(j) (e^{-iH\tau})^j$ and we opt to use GQSP as a natural subroutine for this task. While standard quantum signal processing imposes parity constraints on the achievable polynomials, GQSP enables the direct embedding of an \textit{arbitrary} complex polynomial $P(U)$ of a unitary operator, which is necessary for representing $p_{\omega,\gamma,\tau}(j)$. 

The algorithm employs $U(\tau)$ as the signal operator within the GQSP ansatz. We classically compute $d$ phase angles denoted $\vec{\phi}$ required to synthesize the polynomial $P(z)  =\sum^d_{j=1} c_j z^j  \approx \sum_{j = -j_{\text{max}}}^{j_{\text{max}}}  (\tau / 2\pi) e^{-\gamma \tau |j| + i\omega \tau j}$ approximating the Green's function. The resulting quantum circuit interleaves controlled-$U(\tau)$ operations on the system register with single-qubit rotations $R(\phi_k)$ on an ancilla. The polynomial degree is $d = j_{\text{max}}$, determined strictly by the spectral resolution $\gamma$ required to resolve the feature bandwidth. 

The efficiency of the algorithm is governed by the subnormalization factor $\beta$ in applying $P(z)$ to the linear combination of timestep propagators $U(\tau)$, defined as the 1-norm of the polynomial coefficients. For the finite Fourier series truncation $j_{\text{max}}$, this sum can be derived analytically, starting with the triangle inequality
\begin{equation}
\beta = \sum_{j=-j_{\text{max}}}^{j_{\text{max}}} |p_{\omega,\gamma,\tau}(j)| \leq \frac{\tau}{2\pi} \sum_{j=-j_{\text{max}}}^{j_{\text{max}}} e^{-\gamma \tau |j|}.
\end{equation}
Recognizing this as a geometric series symmetric about $j=0$, we define $r = e^{-\gamma \tau}$ and rewrite the sum as $1 + 2\sum_{j=1}^{j_\text{max}} r^j$. Using the closed form for a geometric progression, we derive the upper bound for $j_\text{max}$ truncated $\beta$ as
\begin{equation}
\beta \leq \frac{\tau}{2\pi} \left( \frac{1 + r - 2r^{j_\text{max} + 1}}{1 - r} \right)
\end{equation}
In the limit of a large truncation cutoff ($j_{\text{max}} \to \infty$), the term $r^{j_\text{max}+1}$ vanishes. Applying the identity $(1+e^{-x})/(1-e^{-x}) = \coth(x/2)$, we recover the standard bound for arbitrary $\beta$ as
\begin{equation}
    \beta \le \frac{\tau}{2\pi} \coth\left(\frac{\gamma \tau}{2}\right) .
\end{equation}
 For the small time steps $\tau$ ($\gamma \tau \ll 1$), we can approximate $\coth(x) \approx 1/x$, yielding the scaling law $\beta \approx 1/(\pi \gamma)$ used in our resource estimates.

Once the GQSP step is completed, the final step is to encompass the subroutines preparing the final state $G_t D_{\mu} \ket{\Psi_I}$ in a Hadamard test. To explicitly quantify the measurement overhead for this entire algorithm, we reconstruct the absolute absorption cross-section $\sigma_A(\omega)$ in the following way. 

First, we classically compute the scalar electric dipole normalization factor $\mathcal{N} = ||D_{\mu} \ket{\Psi_I}||$. In a given molecular orbital basis, evaluating this expectation value on the classical reference state scales polynomially with the number of orbitals, rendering the cost negligible. Second, we perform the aforementioned Hadamard test on the normalized dipole-excited state $|\Phi\rangle = D_{\mu}|\Psi_I\rangle / \mathcal{N}$ utilizing the GQSP unitary $U_\text{GQSP}$ to implement $G_t$. Then, we simply initialize the ancilla in the state $\frac{1}{\sqrt{2}}(|0\rangle - i|1\rangle)$ and measure in the Pauli-$X$ basis to isolate the imaginary component, yielding a random variable $Z \in \{-1, 1\}$. Finally, the estimated physical absorption cross-section is reconstructed as 
\begin{equation}
    \bar{\sigma}_{A}(\omega) = \alpha \cdot \mathcal{N} \cdot \beta \cdot \mathbb{E}[Z],
\end{equation}
 where $\alpha = \frac{4\pi\omega}{3c}$ is the semi-classical prefactor derived from the original absorption cross-section formula in Eq.~\eqref{eq:crosssection}.

The total cost is determined by the number of measurement shots ($M$) required to resolve the target $\sigma_A(\omega)$ to $\epsilon$ accuracy with  $\bar{\sigma}_A(\omega)$. The specific constant factors are governed by three application-specific parameters at the 92 eV operating frequency ($\omega \approx 3.38$ a.u.). Using $c \approx 137$ a.u., the prefactor is 
\begin{equation}
    \alpha \approx 0.10.
\end{equation}
This term effectively suppresses the shot noise when converting to physical units for this specific target photon frequency. Additionally, we fix the spectral broadening to match the experimental bandwidth of EUV sources at 2$\%$ of 92 eV as $\gamma = 1.84$ eV ($0.0676$ a.u.)~\cite{lin2023spectral}, which yields a subnormalization factor of
\begin{equation}
   \beta \approx 1/(\pi \gamma) \approx 4.7 .
\end{equation}
Then, we can estimate the number of shots needed by choosing an error $\epsilon$ in the final absorption cross section value in atomic units. Then the number of shots using the standard Chebyshev bound is
\begin{equation}
    M = (\alpha \mathcal{N}\beta/\epsilon)^2. 
\end{equation} 

\subsection*{Photoemission Spectrum}
\subsubsection*{Background} Upon absorbing an EUV photon, a photoresist monomer can undergo an immediate photoionization event, instead of just a localized absorption, leaving the system in a highly excited non-stationary cationic state characterized by a core-level vacancy. This unstable configuration relaxes primarily through the Auger effect, a correlated two-electron process where a valence electron fills the core hole while transferring energy to a secondary electron, which is subsequently ejected into the ionization continuum.

The primary objective of the simulation presented in this section is to recover the kinetic energy distribution of these ejected electrons, denoted as the \emph{photoemission spectrum}. As mentioned in the introduction, this distribution serves as the essential initial condition for multi-scale photolithography models, parameterizing the initial energies in stochastic electron cascades that eventually determine the blur width of the photoresist.

While the decay rate for a specific transition is formally described by Fermi's golden rule, relying on a perturbative framework to reconstruct the full spectrum is computationally intractable for complex molecules. Fermi's golden rule requires an explicit summation over the exponentially large manifold of final continuum eigenstates to resolve the energy distribution. Furthermore, standard quantum chemistry methods rely on localized Gaussian basis sets, which cannot accurately represent these high-energy, delocalized scattering states.

Therefore, we adopt a non-perturbative, real-time quantum simulation approach to directly simulate the Auger decay process. Instead of summing over individual eigenstates, we simulate the time evolution of the correlated many-body wavefunction in a plane-wave basis using a first quantized representation of the electrons. This naturally generates the superposition of all accessible final states within a chosen simulation cell with a large enough volume to represent the vacuum. We then extract the photoemission spectrum directly by projecting the evolved state onto the continuum subspace and then measuring the kinetic energy in the plane-wave basis. In general, this procedure contains the kinetic energy of both photoionized electrons at short times (sub-fs), and then Auger electrons later in time, but the ionization spectra could be subtracted out from the final time $t$ spectrum to isolate the Auger spectrum specifically. Since this is not a bottleneck, we do not discuss it further in this work for simplicity. Before detailing the full simulation algorithm, we start with a quick overview of first quantized quantum simulations in the next subsection. \\

\subsubsection*{First quantization} In a first-quantized plane-wave representation for electronic structure, the quantum state is encoded in $\eta$ registers corresponding to each electron's momentum, with the total state vector antisymmetrized to satisfy fermionic statistics. This antisymmetrization can be accomplished using the techniques of Ref. \cite{berryImprovedTechniquesPreparing2018}. This algorithm is highly efficient; it prepares the antisymmetrized initial state using $O(\eta n \log(\eta))$ Toffoli gates. Given this represents a small additive prefactor to the overall algorithm, we will forego explicit costing of this step in this work.  For simplicity, using similar notation to Ref.~\cite{su_fault-tolerant_2021}, the system can be defined within a cubic cell of volume $\Omega$, utilizing a basis of $N$ plane waves per electron with reciprocal lattice vectors and periodic boundary conditions
\begin{equation}
\begin{aligned}
    k_p &= \frac{2\pi p}{\Omega^{1/3}} \quad \text{where}\\ \quad p \in G &= \left[-\frac{N^{1/3}-1}{2}, \frac{N^{1/3}-1}{2} \right]^3 \cap \mathbb{Z}^3.
\end{aligned}
\end{equation}

The kinetic energy operator, $T$, is diagonal in the plane-wave basis:
\begin{equation}
T = -\sum_{i=1}^{\eta} \sum_{p \in G} \frac{\|k_p\|^2}{2} |p\rangle\langle p|_i
\end{equation}
where $|p\rangle\langle p|_i$ acts on the register of the $i$-th electron, acting as the identity elsewhere. The electron-nuclear attractive potential, $U$, is
\begin{equation}
U = -\frac{4\pi}{\Omega} \sum_{i=1}^{\eta} \sum_{l=1}^{L} \sum_{p,q \in G, p \neq q} \zeta_l \frac{e^{i k_{q-p} \cdot \mathbf{R}_l}}{\|k_{p-q}\|^2} |p\rangle\langle q|_i
\end{equation}
and the electron-electron repulsion, $V$, couples momentum states via transfer vectors $\nu$:
\begin{equation}
V = \sum_{1 \le i < j \le \eta} \frac{4\pi}{\Omega} \sum_{\substack{p,q \in G \\ \nu \in G_0}} \frac{1}{\|k_\nu\|^2} |p+\nu\rangle\langle p|_i |q-\nu\rangle\langle q|_j.
\end{equation}
The momentum transfer vectors are drawn from the set $G_0 = G \setminus \{(0,0,0)\}$. This representation and Hamiltonian definition will be assumed throughout this section, unless otherwise noted. \\

\subsubsection*{Algorithm}
To target the kinetic energy spectra of photoemitted electrons, one must devise a type of decision boundary in real space for determining if an electron is in the continuum or not. Then, one can compute the kinetic energy spectrum of these ejected electrons after applying this projection. We construct a correlation function for this task of the following form
\begin{equation}
    C(t,\tau) = \bra{\Psi_0} We^{iHt} \Pi_c e^{-iT\tau}\Pi_c e^{-iHt}W\ket{\Psi_0},
    \label{eq:corr-fn-photoemission}
\end{equation}
where $\ket{\Psi_0}$ is the electronic ground state, $W$ is the light matter state preparation operator, including the windowed electric dipole, $t$ is the real simulation time, $\Pi_c$ is a continuum projection operator, $\tau$ is the spectral time length to resolve the signal, and $T$ is the electronic kinetic energy operator. 
One benefit of the plane-wave basis representation of the system we picked earlier is that the correlation function in Eq.~\eqref{eq:corr-fn-photoemission} collapses onto the kinetic energy distribution alone. Since $\Pi_c$ is an orthogonal projector and $T$ is diagonal in the plane-wave basis, inserting the resolution of the identity $\sum_k \ket{k}\!\bra{k} = \mathbbm{1}$ on both sides of $e^{-iT\tau}$ kills the cross terms and leaves
\begin{align}
    C(t,\tau) &= \sum_k e^{-i\varepsilon_k \tau}P_t(k), \nonumber\\
    P_t(k) &:= |\bra{k}\Pi_c e^{-iHt}W\ket{\Psi_0}|^2,
    \label{eq:corr-fn-photoemission-simple}
\end{align}
where $k$ runs over joint momentum configurations of the $\eta$ registers, $\varepsilon_k$ is the total kinetic energy of $k$, and $C(t,0) = \norm{\Pi_c e^{-iHt}W\ket{\Psi_0}}^2$. So $C(t,\cdot)$ is the Fourier transform of $P_t$ over the kinetic energy grid, and carries nothing that $P_t$ does not. Since we can measure the electron's kinetic energy directly in the kinetic energy basis, represented by $\ket{k}$, sampling returns $P_t$ itself, and neither the $e^{-iT\tau}$ evolution nor the transform over $\tau$ has to be built.

To gain some intuition for the correlation function we just defined, we can observe that Eq.~\eqref{eq:corr-fn-photoemission-simple} describes a kind of pump-probe spectroscopy. First, starting from the electronic ground state $\ket{\Psi_0}$, the material is excited with the incident radiation using the windowed dipole $W$ at 92 eV. Then, for a delay time $t$ (typically femtosecond scale for Auger decay, see more in the Application section below), the material undergoes its natural Hamiltonian dynamics. Finally, a probe of incident radiation is then used to measure the kinetic energy of any electrons that are ejected from the material system (modeled by our direct kinetic energy basis measurement with $e^{-iT\tau}$). Our algorithm essentially proceeds by emulating each of these steps.

\begin{figure*}[htp]
    \centering
    \resizebox{\textwidth}{!}{
    \begin{tikzpicture}[
        node distance=1.0cm,
        font=\rmfamily,
        line width=0.7pt,
        draw=black!85,
        wire/.style={
            line width=0.6pt, 
            draw=black!90
        },
        base_box/.style={
            rectangle, 
            rounded corners=2pt, 
            align=center, 
            inner sep=6pt,
            draw=black!80
        },
        style_prep/.style={
            base_box,
            fill=black!3, 
            minimum height=3.5cm, 
            minimum width=1.8cm
        },
        style_w_small/.style={
            base_box,
            fill=prxblue!8, 
            draw=prxblue!60!black, 
            minimum height=1.2cm, 
            minimum width=1.6cm,
            font=\small
        },
        style_w_large/.style={
            base_box,
            fill=prxblue!8,
            draw=prxblue!60!black,
            minimum height=3.5cm, 
            minimum width=1.8cm
        },
        style_time/.style={
            base_box,
            fill=prxred!8, 
            draw=prxred!60!black,
            minimum height=3.5cm, 
            minimum width=1.8cm
        },
        style_proj/.style={
            base_box,
            fill=prxpurple!8, 
            draw=prxpurple!60!black,
            minimum height=3.5cm, 
            minimum width=1.8cm
        },
        measure_box/.style={
            draw=black, 
            fill=white, 
            minimum width=0.8cm,  
            minimum height=0.8cm  
    }
    ]

        \coordinate (top_rail) at (0, 0.8);
        \coordinate (bot_rail) at (0, -0.8);

        \node[anchor=east, font=\large] (ket1) at (-0.6, 0.8) {$|\alpha_1\rangle$};
        \node[anchor=east, font=\large] (ketn) at (-0.6, -0.8) {$|\alpha_\eta\rangle$};
        \node at ($(ket1)!0.5!(ketn)$) {$\vdots$};

        \node[style_prep, right=0.6cm of ket1.east |- {0,0}] (mps) {
            \textsc{mps}\\Orbital\\State Prep.
        };
        \node[style_prep, right=0.8cm of mps] (ferm) {
            Anti-\\symmetrization
        };

        \node[style_w_small] (be_top) at ($(ferm.east) + (2.0, 0.8)$) {\textsc{be}($X$)};
        \node[style_w_small] (be_bot) at ($(ferm.east) + (2.0, -0.8)$) {\textsc{be}($X$)};
        
        \draw[wire] (be_top.south) -- (be_bot.north);

        \node[style_w_large, right=0.6cm of be_top.east |- {0,0}] (qpe) {
            Eigenstate\\Filter
        };

        \node[style_time, right=0.6cm of qpe] (time) {\Large $e^{-iHt}$};

        \node[style_proj, right=0.6cm of time] (proj) {\Large $\Pi_{>R_c}$};

        \node[measure_box] (meas_top) at ($(proj.east) + (0.9, 0.8)$) {};
        \node[measure_box] (meas_bot) at ($(proj.east) + (0.9, -0.8)$) {};

        \foreach \m in {meas_top, meas_bot} {
            \begin{scope}[shift={(\m.center)}]
                \draw[thin] (-0.35, -0.2) arc (150:30:0.4);
                \draw[thick, -{Latex[length=2mm, width=2mm]}] (0, -0.3) -- (0.25, 0.2);
            \end{scope}
        }

        \draw[wire] (ket1.east) -- (mps.west |- top_rail);
        \draw[wire] (mps.east |- top_rail) -- (ferm.west |- top_rail);
        \draw[wire] (ferm.east |- top_rail) -- (be_top.west);
        \draw[wire] (be_top.east) -- (qpe.west |- top_rail);
        \draw[wire] (qpe.east |- top_rail) -- (time.west |- top_rail);
        \draw[wire] (time.east |- top_rail) -- (proj.west |- top_rail);
        \draw[wire] (proj.east |- top_rail) -- (meas_top.west);

        \draw[wire] (ketn.east) -- (mps.west |- bot_rail);
        \draw[wire] (mps.east |- bot_rail) -- (ferm.west |- bot_rail);
        \draw[wire] (ferm.east |- bot_rail) -- (be_bot.west);
        \draw[wire] (be_bot.east) -- (qpe.west |- bot_rail);
        \draw[wire] (qpe.east |- bot_rail) -- (time.west |- bot_rail);
        \draw[wire] (time.east |- bot_rail) -- (proj.west |- bot_rail);
        \draw[wire] (proj.east |- bot_rail) -- (meas_bot.west);

        \begin{scope}[on background layer]
            \node[
                fit=(mps)(ferm), 
                draw=black!60, 
                dashed, 
                dash pattern=on 3pt off 3pt,
                rounded corners=8pt, 
                inner sep=8pt, 
                line width=0.8pt
            ] (group1) {};
        \end{scope}
        \node[below=0.25cm of group1, font=\Large] (lbl_prep) {$|0\rangle \to |\Psi_0\rangle$};

        \begin{scope}[on background layer]
            \node[
                fit=(be_top)(be_bot)(qpe), 
                draw=prxblue, 
                dashed, 
                dash pattern=on 3pt off 3pt,
                rounded corners=8pt, 
                inner sep=8pt, 
                line width=0.8pt
            ] (group2) {};
        \end{scope}
        \node[below=0.25cm of group2, font=\Large] (lbl_w) {$W$};

    \end{tikzpicture}
    }
    \caption{\textbf{Quantum circuit for the photoemission spectrum in first-quantization.} The simulation operates on $\eta$ particle registers initialized as $\ket{\alpha_1}\cdots \ket{\alpha_\eta}$, where each $\alpha$ represents the integer index encoding the momentum coordinates of a single electron. The algorithm proceeds in four main stages: (1) State Preparation: The electronic ground state $\ket{\Psi_0}$ is synthesized from a compressed MPS representation of molecular orbitals, with antisymmetrization applied to enforce fermionic statistics. (2) Excitation ($W$): The electric dipole operator creates the superposition of excited states, using block-encoded position operators ($\textsc{be}(X)$) and Quantum Signal Processing to filter for eigenstates within the 92 eV bandwidth. (3) Time Evolution ($e^{-iHt}$): The wavefunction is propagated for time $t$ to simulate the Auger decay dynamics. (4) Measurement: A real-space projector $\Pi_{>R_c}$ identifies electrons in the continuum. Because the kinetic energy operator is diagonal in the plane-wave basis, the final kinetic energy spectrum is extracted by directly sampling the computational basis of these particle registers. Note that for visual clarity, the extensive ancilla registers required for block encodings and arithmetic subroutines are omitted from this high-level schematic.}
    \label{fig:photoemission_circuit}
\end{figure*}

More concretely, starting from the ground state of the material system, we apply the dipole operator $D_\mu$, which accurately models the material response to broadband radiation in the non-relativistic regime. In this scenario, it is no longer generically classically efficient to apply the dipole operator $D_\mu$ as in the absorption algorithm. This is because the localized molecular orbital basis functions over which we define the dipole operator have arbitrarily small support outside a small radius containing the molecule. This small support means that the transition amplitudes connecting bound and delocalized (continuum) degrees of freedom are artificially suppressed if one tries to prepare the initial state by applying the dipole operator to the molecular orbitals, as we did in the absorption case. Thus, we proceed by applying a block encoding of the dipole operator on the quantum device. The action of $D_\mu$ on $\ket{\Psi_0}$ induces a linear combination of excited states weighted by their transition probabilities as given by Fermi's golden rule. Thanks to the completeness of the plane-wave basis, the dipole transitions in this basis span both bound and continuum states. To emulate the response from a coherent light source with a particular incident frequency, we perform a variant of eigenstate filtering to apply a Gaussian filter function to the spectrum of the material Hamiltonian centered around the $92$ eV central frequency of the radiation. The Gaussian standard deviation parameter $\sigma$ is fixed by the full width at half maximum (FWHM) of the pulse through $\text{FWHM} = {2\sqrt{2\ln(2)}}\,\sigma$. This constructs the $W$ operator of Eq. \eqref{eq:corr-fn-photoemission-simple}. 

Following this, we perform the Hamiltonian dynamics that may induce Auger decay. After time $t$, we then measure the kinetic energy spectrum of the electrons that are delocalized from the material as a result of the Auger decay process. This is obtained by applying a projection operator in real space corresponding to a predetermined cutoff radius distinguishing between those degrees of freedom that are labeled as ``bound'' to and ``unbound'' from the molecular system. The success probability of being in the ``unbound'' subspace gives the probability of observing an electron in the continuum, and sampling the wavefunction in this same subspace provides the kinetic energy spectrum of the continuum wavefunction as a histogram. We summarize these steps in the numbered list below each corresponding to a following subsection:
\begin{enumerate}
    \item Prepare the first quantized ground state from an approximate classical reference in second quantization using the method of Ref. \cite{huggins2025efficient}.
    \item Apply the dipole operator $D_\mu$ to excite the ground state into a linear combination of excited states.
    \item Filter excited states to those that differ from the ground state energy by $\sim 92$ eV.
    \item Perform time evolution of system for $\sim 1$-$10$ fs to allow for Auger decay.
    \item Using a real space boundary, measure kinetic energy of the continuum electronic degrees of freedom.
\end{enumerate}

For a more detailed graphical depiction of the algorithm at a high level, see Fig.~\ref{fig:photoemission_circuit}. \\

\noindent \textit{State Preparation.}\label{sec:state-prep-photoemission}
The first step of our algorithm concerns the preparation of the electronic ground state of the target system upon which the dipole operator will be applied. We assume that for typical molecular systems, and clusters more generally, we start by computing the Hartree-Fock molecular orbitals using a Gaussian basis set, standard in classical quantum chemistry simulation codes. Then, based on this molecular orbital basis, we can choose a single, or multi-Slater determinant representation of the ground state in second quantization, generally described as
\begin{equation}
    \ket{\Psi_0^\text{SQ}} = \sum_{s}^{S} c_s \ket{s}
\end{equation}
where $s$ is the Slater determinant index in second quantization, $S$ is the total number of Slater determinants and $c_s$ is the normalized (real) coefficient. Assuming we have this object precomputed from a chosen classical method, the bottleneck of the state preparation routine is to then transform the compact second quantized ground state, $\ket{\Psi_0^\text{SQ}}$, in the Gaussian orbital basis, to the first quantized representation in the plane-wave basis, $\ket{\Psi_0}$.

We employ one state-of-the-art approach introduced in Ref.~\cite{huggins2025efficient} to prepare molecular ground states in a first-quantized plane-wave basis from a second quantized ground state in a Gaussian basis. This approach circumvents the prohibitive scaling of naive first-quantized plane-wave state preparation, which scales as $O(\eta N)$, by mapping molecular orbitals derived from classical Gaussian basis calculations into a compact Matrix Product State (MPS) form.

The core of the algorithm is the synthesis of a single-particle basis transform unitary, $V$, which maps the computational basis states of the first quantized electron registers to the target MPS-compressed orbitals as a linear combination of plane-wave basis states using standard unitary synthesis techniques. For an $\eta$-electron system, the total non-Clifford gate complexity is dominated by the implementation of this transformation across all registers ($V^{\otimes \eta}$), scaling as
\begin{equation}
    \mathcal{C}_\text{sp} \in \widetilde{O}(\eta K_\text{mo} g^{3/2} \text{polylog}(N)),
\end{equation}
where $K_\text{mo}$ is the number of molecular orbitals, $g$ is the size of the Gaussian basis counting all primitives contained within each orbital basis function, and $N$ is the size of the plane-wave basis. In almost all cases $K_\text{mo}, g \ll N$. Notably, the cost scales only polylogarithmically with the plane-wave basis size, $N$, enabling high-precision discretizations for the state preparation step at relatively minimal cost.

Crucially, the overhead for preparing correlated states beyond a single Slater determinant (e.g., from configuration interaction, DMRG, etc.) is subdominant to preparing the basis transformation operator. This efficiency arises because the computationally expensive basis transformation acts as a fixed dictionary that maps orbital indices to physical wavefunctions; its circuit complexity is determined solely by the number of orbitals in the basis, not by the specific superposition of indices present in the input state. Consequently, once the low-cost initial superposition of integer indices is prepared~\cite{babbush2023quantum}, the dominant cost of synthesizing the physical orbitals remains similar to that of the Hartree-Fock state. Constant factors and polynomial overheads associated with the MPS compression are rigorously minimized and verified using the \texttt{orb2mps-fq} software library of Ref.~\cite{huggins2025efficient}, which demonstrates resource reductions of orders of magnitude for this state preparation task. \\

\noindent\textit{Application of electric dipole operator.} The electric dipole operator is obtained by the dot product of the external electric field, representing the classical light, with the position operator of the system. Let us denote the time dependent external electric field as 
\begin{equation}
\mathbf{E}(t) = \begin{pmatrix}
    E_x(t)\\
    E_y(t)\\
    E_z(t)
\end{pmatrix}
\end{equation}
where $t \geq 0$ and each $E_j(t) \in \mathbb{C}$. Under the dipole approximation, the light-matter interaction is given by the time-independent Hamiltonian
\begin{equation}
    H_{I} = \sum_{j \in \{x,y,z\}}\sum_{i \in [\eta]} \left|E_j\right|e^{i\phi_j} \hat{x}^{(i)}_j,
\end{equation}
where $\hat{x}^{(i)}_j$ is the \textit{position} operator in real space, which acts on spatial direction $j \in \{x,y,z\}$ of particle $i$. Since the electric field is just a constant number ($c$-number) scaling the quantum mechanical operator $\hat{x}^i_j$, its implementation is straightforward. We will focus on the cost to construct a block encoding of a discretization of position operator on a real-space grid. Let $\Omega^{1/3} > 0$ characterize a side length of the ``simulation box'', and discretize the volume $\Omega = [-\Omega^{1/3}/2,\Omega^{1/3}/2]^3$ into $N$ equispaced unit voxels of side $h = \Omega^{1/3}/N^{1/3}$, whose centres are $\mathbf{x} = h\left(a - \tfrac{N^{1/3}-1}{2},\, b - \tfrac{N^{1/3}-1}{2},\, c - \tfrac{N^{1/3}-1}{2}\right)$ with $a, b, c \in \{0,\ldots,N^{1/3}-1\}$. For simplicity, we will always take $N^{1/3} \equiv 2^n$, so that on the quantum computer, the position register of each electron is encoded into exactly $3n$ qubits. We will denote a single electron register as $\ket{\alpha^i} := \ket{\alpha^i_x}\ket{\alpha^i_y}\ket{\alpha^i_z}$, where each $\alpha^i_j \in \{0,1\}^n$ encoding an integer in $[N^{1/3}]$. For $\eta$ electrons, we will write a basis state in the combined system register of $3\eta n$ qubits as $\ket{\alpha} := \ket{\alpha^{0}}\ket{\alpha^{1}}\cdots\ket{\alpha^{\eta-1}}$.

We define the \textit{discrete position operator}, a finite dimensional matrix that is diagonal in the position basis, as
\begin{equation}
\begin{aligned}
    X^i_j\ket{\alpha} &= X\paren{\text{int}\left(\alpha^i_j\right)}\ket{\alpha},
    \\
    X(u) &= \frac{\Omega^{1/3}}{N^{1/3}}\paren{u - \tfrac{N^{1/3}-1}{2}},
    \label{eq:euv-position}
\end{aligned}
\end{equation}
where $\text{int}:\{0,1\}^{n} \rightarrow \{0,\ldots,N^{1/3}-1\}$, so that positions are taken cell-centered. The coordinate origin is a free choice, exact on the $\eta$-electron sector, because a translation by $\mathbf{a}$ sends $X^{(\eta)} \mapsto X^{(\eta)} + \paren{a_x+a_y+a_z}\eta\,\mathbbm{1}$ there. Centering halves the largest coordinate magnitude against a corner origin, and therefore halves the subnormalization factor; it also puts the dipole on the same centered lattice that the continuum projector below uses. Since $X^i_j$ only acts non-trivially on the $j$th dimension register of particle $i$, it takes the form of a one-body operator, embedded into the many-body Hilbert space by tensoring with the identity on all other degrees of freedom.

For both this step, and the time evolution of the system, we will access the position operator and electronic structure Hamiltonian as \textit{block encodings}. A block encoding works by embedding a non-unitary matrix $A$, scaled by a quantity $\lambda$ we call the subnormalization factor ensuring $\norm{A/\lambda} \leq 1$, into the top-left block of a non-unitary matrix $U_A$. Using $m \geq 1$ ancilla qubits, we say that $U_A$ is an $(\lambda, m, \epsilon)$-\textsc{be}($A$) if $\norm{A - I\otimes \bra{0_m}U_A I\otimes \ket{0_m}} \leq \epsilon$.  Block encoding, and the related quantum signal processing (QSP) \cite{lowOptimalQSP,gilyenQuantumSingularValue2019}, allow us to efficiently implement functions of block encoded matrices and in turn is a widely applicable protocol for quantum algorithms. The main primitives to realize the block encodings in this work are the \textsc{select} and \textsc{prepare} routines. Since these routines are common in the quantum algorithms literature, we reserve the definitions of \textsc{select}, \textsc{prepare}, and the block encoding concept more generally, to the Supplementary Information S2-A.

To implement a block encoding of 
\begin{equation}
    X^{(\eta)}:= \sum_{j=1}^{3}\sum_{i=0}^{\eta-1}X_j^i,
\end{equation}
we perform the following operations
\begin{enumerate}
    \item Prepare a uniform superposition over $3\eta$ many degrees of freedom, with non-Clifford cost {$4\ceil{\log(3\eta)}+2b_r-13$}.
    \item Control swaps of a single degree of freedom into an $n$ qubit ancilla register. This requires {$3\eta n + 3\eta - 2$} Toffoli gates.
    \item Perform an $n$ qubit quantum Fourier transform on ancilla register using addition into a phase gradient state {with $b_F$ many qubits. This requires $\mathcal{Q}(n,b_F)$ Toffoli gates}.
    \item Apply $X$ using inequality testing. This requires $n$ Toffoli gates.
    \item Uncompute steps (1-3){, at the cost of steps (1-3)}.
\end{enumerate}
The inequality testing technique used in Step {4}, introduced in Ref. \cite{sandersBlackBox}, is described in the Supplementary Information S2-B. {Neither the transform nor the multiplexed swap is a classical function of a register that survives the routine, so neither can be erased by the ``measure-and-fixup'' construction of Ref.~\cite{gidneyHalvingCostQuantum2018} and both are charged in both directions.} The overall non-Clifford cost to apply the above operations to implement an $({\alpha_X}, {\ceil{\log(3\eta)}+n+1}, \epsilon)$-$\be(X^{(\eta)})$, {at subnormalization}
\begin{equation}
    \alpha_X = 3\eta\,\frac{\Omega^{1/3}}{2},
    \qquad
    \frac{\alpha_X}{\norm{X^{(\eta)}}} = \paren{1-2^{-n}}^{-1},
    \label{eq:euv-dipole-alpha}
\end{equation}
is
\begin{equation}
\begin{aligned}
    &2\paren{4\ceil{\log(3\eta)}+2b_r-13} +\\ &2\paren{3\eta n + 3\eta - 2} + 2\mathcal{Q}(n,b_F) \;+\; n ,
\label{eq:euv-dipole-cost}
\end{aligned}
\end{equation}
where
\begin{equation}
\begin{aligned}
    \mathcal{Q}(n,b) &:= \frac{1}{4}\paren{4nb-2b^2+10b-11}\\
    b_F &= \min\left\{\ceil{\log(n/\varepsilon_F)},\, n\right\},
    \label{eq:euv-qft-toffoli}
\end{aligned}
\end{equation}
{is the phase-gradient approximate transform of Ref.~\cite{park2025reducing} in Toffolis, at gradient width $b_F$ and transform error $\varepsilon_F$. The width is capped at $n$, since an $n$ qubit transform carries no rotation finer than $Z^{-1/2^{n}}$; at $b_F = n$ nothing is discarded and $\mathcal{F}$ is exact. The four terms of Eq.~\eqref{eq:euv-dipole-cost} are the uniform superposition in both directions, the multiplexed swap in both directions, the transform and its inverse, and the comparator.}
{Where measure-and-fixup does apply elsewhere in this work, we remark that} it requires as many ancilla qubits as there are Toffoli gates. However, since the number of ancilla qubits used to block encode the electronic structure Hamiltonian far outnumbers the Toffoli gates (and thereby the number of ancilla) needed to implement the measurement-based uncomputation of these protocols, we may always assume that such an uncomputation can be performed without additional non-Clifford cost. \\

\noindent\textit{Windowed Dipole operator.} To emulate the material response from the interaction with the incident EUV radiation occurring at a specific frequency, we will apply an eigenstate filter to study electronic dynamics restricted to a subspace of energies that could interact with radiation of a particular frequency. The energy of the incident light is 92 eV ($\approx 3.38$ Ha), and therefore cannot induce transitions between eigenstates of the matter Hamiltonian with energy differences significantly far from 92 eV. Typical EUV experiments show that the FWHM of the incident radiation bandwidth is 2\% of the 92 eV energy~\cite{lin2023spectral}, which, for a coherent light source, corresponds to a Gaussian with a standard deviation of $1.84$ eV. In turn, this corresponds to a desired energy resolution of $\delta \equiv 0.067$ Ha and $\delta^{-1} \sim 15$ Ha$^{-1}$.

In order to compute the energy differences required for measuring transition energies, we straightforwardly modify the block encoding of the electronic Hamiltonian by an additive shift of $-\lambda_0 I$, where $\lambda_0$ is the ground state energy estimate of the initial state, in which we can use the computed estimate from the second quantized approach of the classical algorithm piece of the routine in step 1. With this ground state energy value in hand, this shift can be accounted for in the block encoding of $H$ with negligible additional complexity to that of block encoding $H$. This is more efficient than block encoding $H$ and performing a linear combination of block encodings to implement a block encoding of $H - \lambda_0 I$.

To implement the filtering, we begin with this ground-state energy shifted block encoding of the electronic structure Hamiltonian. We will denote the non-Clifford cost to implement the \textsc{select} and \textsc{prepare} portion of the block encoding as $\mathcal{C}_{\textsc{sel}}$ and $\mathcal{C}_{\textsc{prep}}$ respectively: these costs are reported in Eqs. \eqref{eq:prep-costs} and \eqref{eq:sel-cost}. This block encoding is very involved, and the majority of the operations have been described and optimized in great detail in Ref. \cite{su_fault-tolerant_2021}. We follow the notation of this work, and parameters such as $n_M, n_\eta,$ etc. that are introduced below are described in Table S1 of Supplementary Information S3 of this work as well as in Appendix A of the original reference \cite{su_fault-tolerant_2021}. Now, to emulate the response due to the interaction with a coherent light source, we employ a Gaussian filter function on the block encoding. We target a Gaussian centered around $\mu = \frac{3.38}{\lambda}$ Ha with standard deviation $\sigma \frac{\delta}{\lambda}$. Expanding $e^{-x^2/2s^2}$ in Chebyshev polynomials through the scaled modified Bessel functions, only even orders appear, so the parity restriction of quantum signal processing is free, and the truncation error is attained at $x=0$ and equals the discarded tail exactly. Reading that tail as a complementary error function, degree
\begin{equation}
\begin{aligned}
   d_W = \frac{\sqrt{2}\,\text{erfc}^{-1}\!\left(\epsilon\right)\lambda_\omega}{\delta}
\end{aligned}
\label{eq:filter-degree}
\end{equation}
{rounded up to an even integer, suffices, where $\epsilon$ here controls the error at the level of the polynomial approximation. The accuracy therefore enters as $\sqrt{\ln(1/\epsilon)}$ rather than as $\log(1/\epsilon)$, the Gaussian being entire; a rectangular or Lorentzian window would pay the full logarithm at the same $\lambda_\omega/\delta$. In the estimates below we evaluate $\lambda_\omega$ at $\lambda_0 = 0$, adopting no computed ground-state energy.}

The overall non-Clifford cost to perform the quantum signal processing is
\begin{align}
   \mathcal{C}_W &= \mathcal{C}_X + d_W\left(\mathcal{C}_Q + 2(n_T-3)\right) \nonumber \\ &+ (d_W+1)\left(c_{\text{ref}}+b_{\text{pg}}-2\right),
   \label{eq:filter-cost}
\end{align}
{where $\mathcal{C}_Q = \mathcal{C}_{\textsc{prep}} + \mathcal{C}_{\textsc{sel}}$ is one full query $\textsc{prep}^\dagger\textsc{sel}\,\textsc{prep}$, both directions of the preparation already charged inside the itemized terms below, and} $c_{\text{ref}}  = n_{\eta \zeta} + 2 n_\eta + 6n + n_M + 16$ is the number of ancilla qubits reflected against in the block encoding. {The $2(n_T-3)$ is the second rotation that the ground-state shift adds to $\textsc{prep}$, and the last term is the $d_W+1$ phased reflections and synthesized angles that a degree $d_W$ signal-processing sequence carries beside its $d_W$ queries, at a phase gradient of $b_{\text{pg}} = \max\{n_R+1, n_T\}$ bits.} The gate complexity of the preparation routine and selection routine is
\begin{widetext}
\begin{equation}
\begin{aligned}
\mathcal{C}_{\textsc{prep}} &= \underset{\textsc{prep}_{UV}}{\underbrace{\mathcal{A}\left[3n^2 + 15n - 2 + 4n_M(n+1)\right] + 2(2n + 2b_r - 7) + \lambda_\zeta + \min_{k}\left(\ceil{2^{-k} \lambda_\zeta} + 2^k\right)} }\\&
    + \underset{\textsc{prep}_{TUV}}{\underbrace{2(n_T + 4n_{\eta \zeta} + 2b_r -12) + 14n_\eta + 8b_r -36 + 18}}
    \label{eq:prep-costs}
\end{aligned}
\end{equation}
\begin{equation}
\mathcal{C}_{\textsc{sel}} = \underset{\textsc{sel}_{UV}}{\underbrace{24n + C_R}} + \underset{\textsc{sel}_T}{\underbrace{5(n-1)+2}}+ \underset{\textsc{sel}_{TUV}}{\underbrace{12\eta n +4\eta -8}},
\label{eq:sel-cost}
\end{equation}
\end{widetext}
where $\textsc{prep}_{UV}$ (resp. $\textsc{sel}_{UV}$) characterizes the cost to implement the prepare circuit for the potential terms and $\textsc{prep}_{TUV}$ the cost of the prepare circuit that combines $T$ and $U+V$ into the prepare circuit for $H${, with the kinetic preparation and the selector logic folded into the latter. Here $n_\eta = \ceil{\log\eta}$, $\mathcal{A} = 3$ with the amplification round in the momentum-state preparation and $\mathcal{A}=1$ without, and}
\begin{equation}
    C_R = \begin{cases} 3\left(2nn_R - n(n+1) - 1\right), & n_R > n,\\ 3n_R(n_R-1), & n_R \leq n,\end{cases}
    \label{eq:phase-R}
\end{equation}
{is the cost of phasing by $-e^{-ik_\nu\cdot \mathbf{R}_\ell}$. } The parameters $n_R$, $n_T$, $n_M$, and $b_r$ are described in Table S1 of Supplementary Information S3.

A consequence of performing these non-unitary operations is that we incur a success probability overhead. At the level of applying the dipole operator, supposing we have some success probability $P_d$, we would need to apply the quantum circuit for the dipole operator and state preparation $O(1/\sqrt{P_d})$ using amplitude amplification. Then, for the filtering, supposing we have some probability of $P_w$ of successfully applying the filter function, we would need to repeat the quantum circuit that applied the filter and the quantum circuit that applied the dipole $O(1/\sqrt{P_w})$ many times. Overall, including the success probability, the number of gates needed to perform this step is 
\begin{equation}
    \mathcal{C}_{\text{filter}} = O\left(\frac{1}{\sqrt{P_w}}\left(\mathcal{C}_W + \frac{1}{\sqrt{P_d}}(\mathcal{C}_X +\mathcal{C}_{\text{sp}}) \right)\right).
\end{equation}
\\

\noindent\textit{Time Evolution.} This is by far the costliest portion of our algorithm, but is also the most well-developed subroutine we use in this work. We use a quantum signal processing (QSP) based approach to implement the time evolution operator \cite{lowOptimalQSP, Low2019hamiltonian, motlagh2024generalized}, and the cost can be straightforwardly obtained by taking the block encoding costs of Eq. \eqref{eq:prep-costs} and \eqref{eq:sel-cost} and the polynomial degree required to approximate the function $e^{i xt}$. Truncating the Jacobi-Anger expansion, at or above
\begin{equation}
    d \geq \frac{e}{2}\lambda t + \ln\frac{2\eta_J}{\epsilon}, \qquad \eta_J = \frac{4}{\sqrt{2\pi}\,e^{1/13}},
    \label{eq:poly-deg-exp(iH)}
\end{equation}
suffices, where $\lambda$ is the block encoding subnormalization factor, and $\epsilon$ controls the error in the implementation of $e^{iHt}$ at the level of the polynomial approximation. This $\epsilon$ does not control, for example, the error in the block encoding of $H$ itself. Of course, the overall error in $e^{iHt}$ can be obtained by ensuring that $H$ is implemented as a block encoding with error $\epsilon/2$ and $e^{iHt}$ is approximated with error  $\epsilon/2$ and applying a triangle inequality.

Then, applying Eq.~\eqref{eq:poly-deg-exp(iH)}, and the block encoding costs in Eqs.~\eqref{eq:prep-costs} and \eqref{eq:sel-cost}, we obtain a non-Clifford gate cost of
\begin{equation}
\begin{aligned}
    \mathcal{C}_\text{te}& = d\left(\mathcal{C}_Q + c_\text{ref}+1\right) + 2\left(\mathcal{C}_Q c_\text{ref}-1\right) \\ &+   (2d+3)(b_{\text{pg}}-3)
\end{aligned}
\end{equation}
to perform the time evolution up to time $t$. The first term is the $d$ directional walk steps, each a query plus a reflection on the $c_\text{ref}$ ancillas by unary iteration; the second is the two bracketing steps that select the walk against the identity, which is what an indefinite-parity target costs over a definite-parity one; and the third is the $2d+3$ single-qubit rotations synthesized against the phase gradient.

\begin{table*}[ht]
    \centering
\begin{tblr}{
  colspec = {|c|c|Q[c,wd=6cm]|},
  vlines,
  hlines,
  }
\SetCell[c=3]{c} \textbf{Photoemission Spectrum Algorithm Subroutines} \\
\textbf{Symbol} & \textbf{Cost} & \textbf{Description} \\
$\mathcal{C}_\text{sp}$ & $\widetilde{O}(\eta K_\text{mo} g^{3/2} \text{polylog}(N))$ & Initial state preparation (Constant factors computed via \texttt{orb2mps-fq} library of Ref.~\cite{huggins2025efficient}) \\
$\mathcal{C}_X$ & {Eq. \eqref{eq:euv-dipole-cost}} & Block encoding of position operator used to apply dipole operator \\
$\mathcal{C}_\textsc{prep}$ & Eq. \eqref{eq:prep-costs} & Cost to implement \textsc{prepare} circuit in block encoding of $H$ \\
$\mathcal{C}_\textsc{sel}$ & Eq. \eqref{eq:sel-cost} & Cost to implement \textsc{select} in block encoding of $H$ \\
$\mathcal{C}_{\textsc{bound}}$ & {$\eta\left(6\mathcal{Q}(n,b_F) + 3n^2 - n - 1\right) + \eta - 1$} & Projector onto ``continuum'' real-space degrees of freedom \\
$\mathcal{C}_W$ & {Eq. \eqref{eq:filter-cost}, at $d_W = \frac{\sqrt{2}\,\text{erfc}^{-1}\!\left(\epsilon\right)\lambda_\omega}{\delta}$} & Gaussian eigenstate filter from dipole excited ground state\\
$\mathcal{C}_\text{te}$ & {$d\left(\mathcal{C}_Q + c_{\text{ref}}+1\right) + 2\left(\mathcal{C}_Q+c_{\text{ref}}-1\right) + (2d+3)(b_{\text{pg}}-3)$} & Time evolution of the system Hamiltonian with normalization factor $\lambda$ for some time $t${, at the degree $d$ of Eq. \eqref{eq:poly-deg-exp(iH)}}.
\end{tblr}
\caption{\textbf{Costs of photoemission algorithm subroutines.} Here we report the non-Clifford gate costs of the main subroutines used in the photoemission simulation algorithm.}\label{tab:photoionization-costs}
\end{table*}

\noindent\textit{Measurement.} Finally, our goal is to extract the distribution of kinetic energies of electrons that are emitted from the molecule into the continuum. We use a sphere centered at the origin in real space (i.e., at $(0,0,0)$) with a fixed radius, denoted $R_c$, to bifurcate the configuration space into ``bound'' and ``continuum''. This provides a quantitative decision boundary; any electronic degree of freedom with real-space coordinate $x$ satisfying $\norm{x} \geq R_c$, belonging to the ``continuum'', otherwise belonging to the localized molecular bound states. We choose a hard cutoff for simplicity, but soft boundary extensions would be straightforward, similar to classical simulation methods applied to small models of photoelectron emission in the literature~\cite{de2012ab}. The choice of $R_c$ is dependent on the molecular system and the energy of the initial state. For now, we will assume that $R_c$ is a given fixed constant.

Let us recall the lattice of points
\begin{equation}
    q \in \left[-\frac{N^{1/3}-1}{2},\frac{N^{1/3}-1}{2}\right]^3 \equiv G \subset \mathbb{Z}^3,
\end{equation}
where a point in the real space is given by 
\begin{equation}
    r_{q} = \frac{\Omega^{1/3}}{N^{1/3}}{q}.
\end{equation}
Applying the inverse quantum Fourier transform to a point $\ket{{p}}$ in the reciprocal space yields a superposition of points ${q}$ in the real space. Each point ${q} \in G,$ is associated with the tuple of integers $(q_x, q_y, q_z)$, with each $q_j$ a binary value encoding the subset of integers $\left[-\frac{N^{1/3}-1}{2}, \frac{N^{1/3}-1}{2}\right]\cap \mathbb{Z}$. The element is mapped to the associated position in the discretized real space by the scaling factor $\frac{\Omega^{1/3}}{N^{1/3}}(q_x,q_y, q_z)$. Using this relation, we can determine if ${q} \in G$ lies inside a sphere of radius $R_c > 0$ by checking the inequality,
\begin{equation}
    q_x^2 + q_y^2 + q_z^2 <  \left(\frac{R_c N^{1/3}}{\Omega^{1/3}}\right)^2.
    \label{eq:rc-ineq}
\end{equation}

Notice that in contrast to the position operator, we need to test if the norm of a vector in 3 spatial dimensions satisfies the inequality \eqref{eq:rc-ineq}, {so all three axis registers of an electron must be transformed rather than one. The transform on a cubic lattice is a tensor product across the axes, so this is three $n$ qubit transforms per electron and not a transform on $3n$ qubits. The comparison needs no separate comparator either: initializing a $(2n+3)$ bit accumulator to $2^{2n+2} - \floor{(R_c N^{1/3}/\Omega^{1/3})^2}$ by $X$ gates and summing the three squares into it leaves the test of Eq.~\eqref{eq:rc-ineq} in its most significant bit.} The overall costs to perform each subroutine to realize this operation for a single electron are
\begin{enumerate}
    \item Quantum Fourier transform to switch into the position basis{, and its inverse. This requires $6\mathcal{Q}(n,b_F)$ Toffoli gates}.
    \item Summation of three squares {into the offset accumulator, whose most significant bit is the test of Eq.~\eqref{eq:rc-ineq}}. This requires $3n^2 - n - 1$ Toffoli gates.
    \item {Uncompute (2), and then (1).}
\end{enumerate}
{The accumulation retains its carries, so step 2 is erased by its Clifford inverse at no Toffoli cost, whereas the transform is charged in both directions for the reason given for the dipole above, which is the factor of six in step 1. This leaves a total Toffoli cost of $6\mathcal{Q}(n,b_F) + 3n^2 - n - 1$ per electron, independent of $R_c$ entirely, since the comparand is classically precomputed and enters only through the $X$ layer that loads it.} We combine the above steps into a single subroutine we denote as $\textsc{bound}$, which performs the operation
\begin{equation}
    \ket{{q}}\ket{0} \xrightarrow{\textsc{bound}} \ket{{q}}\ket{b(\norm{q} < R_c)},
    \label{eq:?-bound-1body}
\end{equation}
where $b(\text{false}) = 0$ and $b(\text{true})=1$. This notation neglects explicitly writing out the ancilla registers that are used and uncomputed along the way, as described above. 

For the many-body case, the decision boundary becomes if there is any single electron register in the continuum. To test for this, we can simply append an ancilla register of $n_\eta = \ceil{\log(\eta)}$ qubits, then perform steps 1-3 above, and controlling on the $\ket{0}$ state of the flag qubit $\ket{b(\norm{q_i}<R_c)}$ in Eq. \eqref{eq:?-bound-1body}, perform controlled addition of $+1$ into the $n_\eta$ qubit ancilla register and uncompute steps 1-3. Repeating this operation $\eta$ many times, and with a flag qubit set to the $\ket{0}$ state, we control on the all-zero state of the $n_\eta$ qubit register to apply an $X$ gate to the flag. Depending on the qubit budget, it is also possible to remove the costs associated with performing controlled addition by $1$ by using $\eta$ ancilla qubits to flag each register as bound $\ket{0}$ or continuum $\ket{1}$, by combining the $\eta$ flag qubits onto a single ancilla by a binary tree of temporary logical-\textsc{and} gates. This results in a Toffoli cost of $\eta-1$ as opposed to the $\eta \log(\eta)$ cost in the former approach. We take the second route and retain the $\eta$ flags rather than combining them away, which costs nothing: the tree is erased by measurement-based uncomputation regardless, and the flags are live only while the projector runs, below the peak that the propagator sets.  In either case, we may combine the above steps into the operator \textsc{all-bound}, which performs the operation
\begin{equation}
    \ket{q^{(\eta)}}\ket{0}\xrightarrow{\textsc{all-bound}} \ket{q^{(\eta)}}\ket{\prod_{i=0}^{\eta -1}b(\norm{q_i} < R_c)},
    \label{eq:all-bound}
\end{equation}
where we introduced the shorthand notation $\ket{q^{(\eta)}} \equiv \ket{q_0,q_1,\ldots,q_{\eta-1}}$.

Applying this operation to a superposition performs a projection
\begin{equation}
\begin{aligned}
   \ket{\psi} &=\sum_{{q^{(\eta)}}}\psi(q^{(\eta)})\ket{q^{(\eta)}}\ket{0}\\
    \xrightarrow{\textsc{all-bound}}\\
    &\sum_{q^{(\eta)}: \exists \norm{q_i} \geq R_c } \psi(q^{(\eta)})\ket{q^{(\eta)}}\ket{0} +\\& \sum_{q^{(\eta)} :\nexists \norm{q_i} \geq R_c}\psi(q^{(\eta)})\ket{q^{(\eta)}}\ket{1} \\
    &= \Pi_{\text{cont}} \ket{\psi} \ket{0} + \Pi_{\perp}\ket{\psi}\ket{1},
\end{aligned}
\end{equation}
where $\Pi_{\text{cont}}$ is the projector onto the subset of the many-body wavefunction where \textit{any} particle has a position coordinate with radius larger than or equal to $R_c$, and $\Pi_{\perp}$ the projector onto the set of states where \textit{all} particle position coordinates have radius less than $R_c$. Taking the second of the two routes above, the operation $\textsc{all-bound}$ is $\eta$ times the cost for a single particle plus the $\eta-1$ Toffoli gates of a binary tree combining the $\eta$ flags, and can therefore be realized with non-Clifford cost 
\begin{equation}
\begin{aligned}
\label{eq:cost-real-space-proj}
   \mathcal{C}_{\textsc{bound}}&= \eta\left(6\mathcal{Q}(n,b_F) + 3n^2 - n - 1\right) + \eta - 1.
\end{aligned}
\end{equation}

As before, application of a non-unitary operation will incur some success probability overhead. Were the algorithm to fail at this stage, we would need to reflect against all the other steps of the algorithm, leading to a cost of 
\begin{equation}
    \mathcal{C}_{\text{tot}} = \frac{1}{\sqrt{P_{c}}}\left(\mathcal{C}_{\textsc{bound}}+\mathcal{C}_{\text{te}} + \mathcal{C}_{\text{filter}}\right),
\end{equation}

which would be catastrophic if $P_c$ were small. {Writing $\Pi^{(i)}_{<R_c}$ for the projector onto configurations with electron $i$ inside the cutoff, so that $\Pi_\perp = \prod_i \Pi^{(i)}_{<R_c}$, these $\eta$ single-particle projectors are diagonal in the position basis and commute. Hence $\Pi_\perp \leq \Pi^{(i)}_{<R_c}$ for every $i$, and}
\begin{align}\label{eq:pc-bound}
    P_c \geq & 1 - \min_i \bra{\psi}\Pi^{(i)}_{<R_c}\ket{\psi}, \\
    & \text{where} \quad \ket{\psi} = \frac{e^{-iHt}W\ket{\Psi_0}}{\norm{e^{-iHt}W\ket{\Psi_0}}}, \nonumber
\end{align}
{the continuum weight of the most ionized electron. That quantity is the continuum yield at the 92 eV line: the probability that the photon ejects an electron past $R_c$ within the Auger window. It is a property of the resist, and is itself one of the quantities the calculation reports, so we carry it as a parameter of the instance rather than fixing it by an assumption. The circuit does not change with it; only the number of amplification rounds does, and the total scales as $P_c^{-1/2}$. The estimates below are reported at unit yield, and any other value multiplies them by $P_c^{-1/2}$.}

Now by applying the operations that we discussed above we have created the state
\begin{equation}
    \Pi_{\text{cont}} e^{-iHt} W \ket{\Psi_0},
\end{equation}
omitting normalization factors. Now, we can simply sample from this post-selected state to estimate the distribution of single-particle kinetic energies of the continuum electronic degrees of freedom. Each shot returns the flag pattern alongside the momenta, and the flagged registers are the continuum electrons. From each particle register $i$ we measure and obtain a momentum vector $k_i$. We then classically compute the quantity $\frac{\norm{k_i}^2}{2}$. Writing $F = \{i : b_i = 1\}$ for the flagged set, which is non-empty on the success branch, each shot supplies $\lvert F\rvert$ samples of the continuum electrons' kinetic energy distribution, and we define the joint statistic
\begin{equation}
    T^{(k)} = \frac{1}{\lvert F\rvert}\sum_{i \in F} T_i^{(k)},
\end{equation}
where
\begin{align*}
    T_i^{(k)} = \begin{cases}
        1 & \text{register } i \text{ has kinetic energy $k$}\\
        0 & \text{else}.
    \end{cases}
\end{align*}
Since each $T_i^{(k)}$ is binary, the sampling-based estimator is a Bernoulli process with probability $p = \mathbb{E}[T^{(k)}]$, and variance $\sigma^2 = p(1-p) \leq \frac{1}{4}$.

Its mean is the kinetic energy density of the continuum electrons themselves, which is the photoemission spectrum sought. The $\Pi^{(i)}_{<R_c}$ are position-diagonal and commute, so the pattern and the momenta are measured in sequence exactly as the single flag and the momenta are. Auger decay ejects a photoelectron and then an Auger electron, so $\lvert F\rvert$ is typically two within the window; we set $\lvert F\rvert = 1$ below.

For a fixed number of samples $M$, obtaining samples $T_m^{(k)}$, we have the estimator
\begin{equation}
    \hat{p}(k) = \frac{1}{M}\sum_{j=1}^{M}T_j^{(k)}.
\end{equation}
The variance is,
\begin{align*}
    \text{Var}\left(\hat{p}(k)\right) = \frac{\text{Var}\left(T^{(k)}\right)}{M} = \frac{\sigma^2\left(1+({\lvert F\rvert}-1)\rho\right)}{{\lvert F\rvert} M},
\end{align*}
where the last equality results from the fact that for a linear combination of statistics $T' = \sum_{i=1}^{\eta}a_i T'_i$ we have 
\begin{align*}
    \text{Var}\left(T'\right) = \left[\sum_{i=1}^{\eta}{a_i^2}\,\text{Var}\left(T'_i\right) + \sum_{i\neq j} a_i a_j\text{Cov}\left(T'_i,T'_j\right) \right].
\end{align*}
Then, with $a_i = a_j = {\frac{1}{\lvert F\rvert}}$, for identical particles $\text{Cov}(T_i,T_j) = \rho \sigma^2$ for all $i,j$ and $\text{Var}(T_i) = \sigma^2$ for all $i$, we obtained the desired statement.

Now, to bound the sampling complexity, we can apply Bernstein's inequality
\begin{align}
    \mathbb{P}&\left( \left|\frac{1}{M}\sum_{j=1}^M T^{(k)}_j - \mathbb{E}[T^{(k)}] \right| \geq \epsilon \right) \nonumber \\ &\leq 2 \exp\left(-\frac{M\epsilon^2}{2 v + \frac{2}{3}\epsilon}\right),
\end{align}
the samples $T_j$, with variance upper bounded by $v$. {Inverting it,}
\begin{equation}
    M \geq \frac{2\sigma^2}{\epsilon_p^2}\ln\frac{2}{\delta_M} + \frac{4}{3\epsilon_p}\ln\frac{2}{\delta_M}
    \label{eq:shot-count-flagged}
\end{equation}
{samples give $\lvert\hat{p}(k) - p(k)\rvert \leq \epsilon_p$ at a given kinetic energy $k$ with probability at least $1-\delta_M$; a simultaneous guarantee over all $\lvert\mathcal{E}\rvert$ outcomes replaces $\ln(2/\delta_M)$ by $\ln(2\lvert\mathcal{E}\rvert/\delta_M)$. At $\epsilon_p = 10^{-2}$ and $\delta_M = 0.05$ this is $M = 1.89\times10^{4}$, independent of the grid and of $\eta$. These are flagged shots, so the physical count is larger by the reciprocal of the success probability without amplification.}

This completes the characterization of the cost to perform the photoemission spectrum algorithm. The cost for each subroutine is provided in \Cref{tab:photoionization-costs}. \\

\subsection*{Application: IMePh Resource Estimation}
\subsubsection*{Model system} 
\begin{figure}
    \centering
    \includegraphics[width=0.5\linewidth]{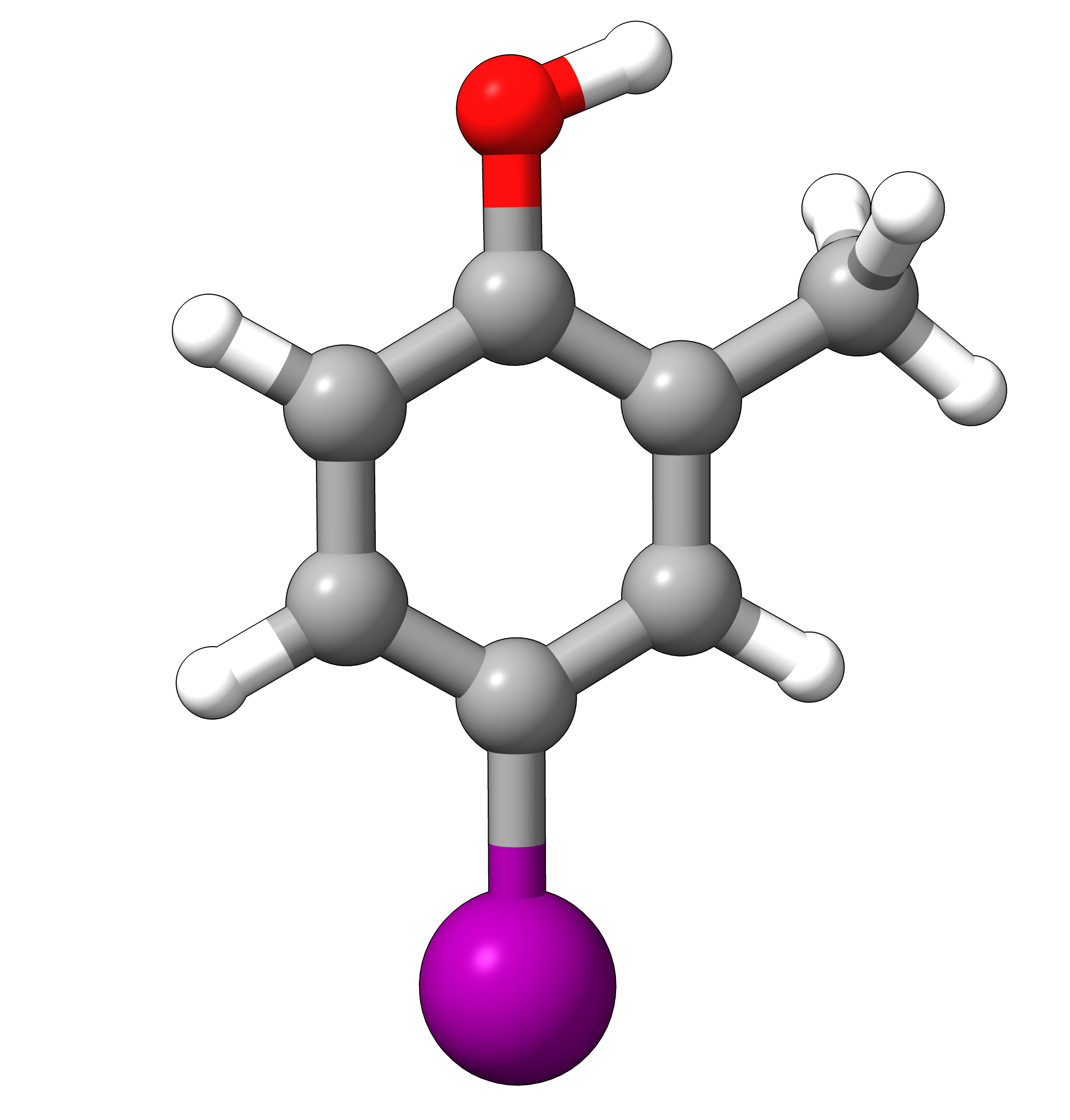}
    \caption{\textbf{Molecular structure of 4-iodo-2-methylphenol (IMePh).}~\cite{PubChem2025_CID143713} White represents hydrogen atoms, grey for carbon, red for oxygen, and purple for iodine. For an all-electron representation of this system, $\eta = 110$. }
    \label{fig:imeph}
\end{figure}

In order to provide concrete estimates of the quantum simulation costs on fault tolerant quantum computers for the two physical quantities introduced above for EUV photolithography, we model a single molecule of 4-iodo-2-methylphenol as a representative monomer of a larger photoresist material, as shown in~\Cref{fig:imeph}. The inclusion of iodine is critical for resist performance, as its large cross-section at 92 eV dominates photon absorption compared to other halogenated materials~\cite{closser2017importance}. By focusing on this single molecule model, we isolate the essential physics, specifically the excitation of iodine 4$d$ core states and the subsequent Auger decay cascade, that generates the low-energy electrons responsible for secondary electron effects. This reduction to a molecular Hamiltonian is physically justified by the ultrafast, localized nature of core-level relaxation, allowing us to tractably estimate the electronic dynamics while bypassing the computational complexity of simulating large bulk amorphous polymers. \\

\subsubsection*{Absorption Sensitivity}\label{ssec:application-absorption}
First, we provide quantum resources estimates for computing the first quantity, absorption sensitivity of IMePh at 92 eV. We only consider logical qubit count and non-Clifford gate complexity throughout our resource estimation unless otherwise noted. A summary of the resource estimates can be found in~\Cref{tab:imeph_absorption_costs}. 
\begin{table*}[t]
    \centering
    
    \begin{tabular*}{\textwidth}{@{\extracolsep{\fill}} l S[table-format=3.0] S[table-format=1.2e2] S[table-format=1.2e2] }
        \toprule
        \multirow{2}{*}{\textbf{Number of Orbitals}} & \multicolumn{3}{c}{\textbf{Resource Estimates}} \\ 
        \cmidrule(lr){2-4} 
        & {Logical qubits} & {Toffoli gates per circuit} & {Overall Toffoli count} \\
        \midrule
        22 & 148  & 3.94e9  & 3.40e12 \\
        28 & 160  & 8.14e9  & 7.02e12 \\
        34 & 172  & 1.46e10 & 1.26e13 \\
        40 & 184  & 2.38e10 & 2.05e13 \\
        50 & 204  & 4.65e10 & 4.01e13 \\
        \bottomrule
    \end{tabular*}
    \caption{\textbf{Absorption Sensitivity Resource Estimation.} The table lists the required number of logical qubits and maximum number of Toffoli gates per circuit, and the total for varying active space sizes for computing the absorption sensitivity of IMePh at 92 eV using the coherent-time absorption algorithm for the IMePh molecule in the def2-SVP basis. The shots overhead that is accounted for in the ``Overall Toffoli count'' column is obtained using the standard Chebyshev bound.}
    \label{tab:imeph_absorption_costs}
\end{table*}
The construction of the model Hamiltonian and the application  is described in the Methods section. From these quantum resource estimations, we see that resolving the absorption sensitivity at a single frequency such as 92 eV is a low-medium cost quantum algorithm, with low logical qubit overhead. The maximum circuit sizes falling within the threshold of the capabilities of expected utility-scale fault-tolerant devices, but further optimizations will need to be made to make it viable for larger target monomers, such as tin-oxide materials. \\

\begin{table*}[t]
    \centering
    \setlength{\tabcolsep}{10pt}
    \begin{threeparttable}
    \begin{tabular}{
        l                      
        c                      
        S[table-format=3.0]    
        S[table-format=4.0]    
        S[table-format=1.2e2]  
        S[table-format=1.2e2]  
    }
        \toprule
        \multicolumn{3}{c}{\textbf{Simulation Parameters}} & \multicolumn{3}{c}{\textbf{Resource Estimates}} \\
        \cmidrule(r){1-3}\cmidrule(l){4-6}
        {Method} & {Basis Size} & {Time (fs)} & {Qubits} & {Gate Cost} & {Overall Cost} \\
        \midrule

        \multirow{2}{*}{AE} & \multirow{2}{*}{$N_\text{pw}^{1/3}=2^9$}
                & 1  & {{5027}} & 7.77e14 & 1.47e19 \\
           &    & 10 & {{5028}} & 4.52e15 & 8.56e19 \\
        \addlinespace

        \multirow{2}{*}{AE} & \multirow{2}{*}{$N_\text{pw}^{1/3}=2^{11}$}
           & 1  & {{6217}} & 4.39e15  & 8.31e19 \\
           &    & 10 & {{6217}} & 2.55e16 & 4.84e20 \\
        \addlinespace

        \multirow{2}{*}{AE} & \multirow{2}{*}{$N_\text{pw}^{1/3}=2^{13}$}
           & 1  & {{7463}} & 3.12e16 & 5.91e20 \\
           &    & 10 & {{7463}} & 1.82e17 & 3.44e21 \\
        \addlinespace

        \multirow{2}{*}{PP} & \multirow{2}{*}{$N_\text{pw}^{1/3}=2^6$}
           & 1  & {{2233}} & 4.67e14 & 8.85e18 \\
           &    & 10 & {{2233}} & 2.72e15 & 5.15e19 \\
        \addlinespace
        
        \multirow{2}{*}{PP} & \multirow{2}{*}{$N_\text{pw}^{1/3}=2^8$}
           & 1  & {{3193}} & 2.98e16 & 5.65e20 \\
           &    & 10 & {{3193}} & 1.74e17 & 3.29e21 \\
        \addlinespace
        
        \multirow{2}{*}{PP} & \multirow{2}{*}{$N_\text{pw}^{1/3}=2^9$}
           & 1  & {{3552}} & 4.13e17 & 7.83e21 \\
           &    & 10 & {{3552}} & 2.41e18 & 4.56e22 \\
        \bottomrule
    \end{tabular}
    \end{threeparttable}
    \caption{\textbf{Photoemission Spectrum Resource Estimates.}
    The table lists the required logical qubits and non-Clifford gate counts for varying the plane-wave grid size and simulation times for the first-quantized photoemission algorithm. We compare both the all electron (AE) method and pseudopotential (PP). These numbers use a computational volume of {$\Omega = 200^3\,a_0^3$}, and are reported assuming that the success probability of applying the dipole operator and window function $P_d=10^{-3}=P_w$, and that $P_{c} =1${, at $\epsilon = 1.6\times10^{-3}$ on each use of the block encoding and $\epsilon_W = 10^{-3}$ in the filter}. The gate cost reflects the number of non-Clifford gates in any individual circuit, and the overall cost includes the sampling complexity of repeating each circuit {the $M = 1.89\times10^{4}$ times of Eq.~\eqref{eq:shot-count-flagged} at $\epsilon_p = 0.01$ and $\delta_M = 0.05$, the same for every row}. The number of plane-wave modes for AE and PP is reflective of the fidelity each method provides in terms of representing the ground state wavefunction in a second quantized basis. The first AE and PP rows show the cost to provide $> 10\%$ fidelity, the second $> 70 \%$ fidelity, and the final rows to $> 99\%$ fidelity with the second quantized ground state. We observe that in all instances PP uses {roughly half the logical qubits, and that its gate cost crosses over: $0.60$ times the AE cost at the first pairing, $6.8$ times at the second and $13.2$ at the third, since the pseudopotential block encoding carries a term that grows with the number of plane waves}.}
    \label{tab:auger_costs}
\end{table*}

\subsubsection*{Photoemission Spectra}
Second, we compute resource estimates for the photoemission simulation algorithm for IMePh, using the first quantized photoemission algorithm. Similar to the previous subsection, we report logical qubit counts, as well the total number of non-Clifford gates to compute the target quantity. A summary of the resource estimates can be found in~\Cref{tab:auger_costs} for the all-electron system. Additionally, for the following results, we assume that we can reduce the number of electrons in the simulation from the all electron approach at $\eta=110$, down to $\eta = 58$, using pseudopotential based methods. This reduction in electron count is detailed in the Supplementary Information S3. Pseudopotential based methods slightly modify the electron-nuclear interaction term, freeze out core electrons, and reduce the overall cost of the algorithm. Details of this modification are detailed in the Supplementary Information S4. The costs of both models are reported for completeness. 

From these results, we see that while the real-time evolution of the physical process we are simulating is small ($\sim1-10$ fs) due to the ultrafast nature of Auger decay, the overhead in the gate costs driving up the number of non-Clifford gates is primarily due to the one-norm of the first quantized Hamiltonian in qubitized time evolution. Some of this cost is mitigated by using a pseudopotential approach, reducing the total electrons in the system, but the overheads remain high. Specifically, for this EUV application, large cell size volumes are necessary to capture the continuum or electron emission events properly, versus typical electronic structure calculations. While pseudopotentials can still be beneficial for this simulation, the fact that the real-space distribution of the pseudopotential function is a smaller percentage of the total cell volume limits the size of the cost reduction, and can even be more costly than the all-electron approach in certain cases. At the same time, the logical qubit count is moderate, on the order of $1000$s.

\section*{Discussion}\label{sec:conclusion}

In this work, we have introduced two new quantum simulation algorithms to address critical material design challenges facing Extreme Ultraviolet (EUV) lithography. As the semiconductor industry pushes toward smaller and smaller features on the atomic scale, the line edge roughness and blur driven by secondary electron cascades induced by EUV photons has emerged as a fundamental bottleneck limiting the further reduction of transistor size, and is difficult to computational model accurately. The quantum simulations proposed here and the subsequent observables, serve as high-fidelity parameterization inputs for broader multi-scale models that use molecular dynamics, kinetic Monte Carlo methods, and more. By providing accurate \emph{ab initio} data for the absorption sensitivity and the Auger spectrum of ejected electrons, these quantum simulations enable better calibration of macroscopic exposure models that are otherwise limited by the accuracy of their underlying parameters.

To address these key parameters, we developed two distinct simulation protocols targeting the primary mechanisms of EUV exposure. First, we presented a coherent time-domain algorithm to compute the absorption cross-section specifically at the 92 eV operating frequency. By bypassing full spectrum algorithmic approach in favor of a more targeted dipole autocorrelation function, we demonstrated that candidate monomers like 4-iodo-2-methylphenol (IMePh) can be screened with relatively low quantum resources, estimated at fewer than $ 200$ logical qubits with costs per circuit of $10^{9}$ total non-Clifford gates and approximately $10^3$ shots for the smallest instance. Second, we addressed the complex electron dynamics of Auger decay and photoelectron emission by utilizing a first-quantized plane-wave basis representation. This approach, coupled with a novel dipole energy windowing, and real-space projection operator to distinguish bound versus continuum states, allows for the rigorous sampling of the ejected electron's kinetic energy spectrum.  The total cost of this novel algorithm is currently still high for reasonably sized grid spacings, {$\geq 10^{14}$ total non-Clifford gates per circuit and $10^4$ shots}, but the logical qubit count is moderate, on the order of $10^3$. {Retaining the per-electron continuum flags, rather than combining them into a single bit, is what keeps the shot count at $10^4$: it identifies the continuum registers on each shot, so the histogram is built from the ejected electrons themselves rather than from a one-body density in which they carry a fraction $1/\eta$ of the weight. Discarding that information costs a factor $\eta$ in shots for no saving in gates.}

Future work will focus on optimizing the algorithmic primitives identified as the primary computational bottlenecks in this study. Specifically, the cost of the photoelectron emission simulation is currently dominated by the extensive gate overhead required for Hamiltonian time evolution in the first-quantized basis, though by a margin that depends strongly on the window. Both the propagator degree and the filter degree $d_W$ are proportional to the block-encoding subnormalization, but only the propagator carries $t$: at every grid in~\Cref{tab:auger_costs} it costs $1.15$ times the filter over a $1$ fs window and $11.5$ times over $10$ fs. Reducing the one-norm is what helps both. Research into hybrid basis representations could significantly reduce the circuit depth required to propagate the wavefunction into the ionization continuum. We stress that this work provides high-fidelity molecular-scale inputs for EUV-induced electron cascade multi-scale modeling. Embedding these quantities into bulk condensed-phase environments and consistently coupling them to stochastic electron transport simulations are essential towards quantitatively predicting macroscopic resist blur.

\section*{Methods}

\subsection*{Time evolution for absorption algorithm}
The time evolution unitary used in the absorption algorithm is implemented using the Trotter-based compressed double factorized approach. This entails factorizing the two-electron integrals as a sum of $L$ individual Hamiltonian fragments, each of which takes a diagonal form $Z^{(\ell)}$ in the basis defined by single-particle basis rotations $U^{(\ell)}$ 
\begin{align}\label{eq:(pq|rs)}
   (pq|rs) \approx \sum_{\ell=1}^L \sum_{k,l=1}^N U^{(\ell)}_{pk} U^{(\ell)}_{qk} Z^{(\ell)}_{kl} U^{(\ell)}_{rl} U^{(\ell)}_{sl}.
\end{align}
The one-electron term can similarly be factorized as
\begin{equation}
    (p|\kappa|q) = \sum_k \tilde{U}_{pk}^{(0)} \tilde{Z}_{kk}^{(0)} \tilde{U}_{qk}^{(0)}.
\end{equation}
With the CDF approach, once the Jordan-Wigner mapping is employed, the Hamiltonian can be transformed into a sum of single-qubit and two-qubit Pauli $Z$ rotations, interspersed with unitaries $\bm{U}^{(\ell)}$ that rotate many-body wavefunctions in the full Hilbert space in response to changing the single-particle basis by a $U^{(\ell)}$ rotation. These unitaries may be synthesized using Thouless' theorem
\cite{kivlichan2018quantum,thouless1960stability},
\begin{align}
    \bm{U}^{(\ell)} = \exp\left(\sum_{p,q}[\log U^{(\ell)}]_{pq} (a^\dagger_p a_q -a^\dagger_q a_p) \right),
\end{align}
and subsequently implemented by decomposing it into Givens rotations \cite{arrazola2022universal}.

The different fragments are diagonal in separate bases and as such do not commute with each other, nor with the fragment from the 1-electron Hamiltonian term. Given this, we implement the overall time evolution unitary by treating the different fragments with the second-order Trotter product formula. This incurs a Trotter error, which results in shifts of the peaks in the computed spectrum. We estimate this shift using perturbation theory \cite{mehendale2023estimating,fomichev2025fast}, leveraging directly the estimates made in Ref. \cite{fomichev2025fast} for the magnitude of Trotter error of similar-sized systems and its extrapolated scaling growth. Combining the Trotter error estimates with the per-step cost of implementing the single- and two-qubit Pauli $Z$ rotations and unitaries $\bm{U}^{(\ell)}$ allows to directly estimate the constant-factor resources required to implement $e^{-iH\tau }$: the process follows closely the approach in Ref. \cite{fomichev2025fast} (additional details are provided in the Supplementary Information S1). \\

\subsection*{Model IMePh Hamiltonian for absorption}

\noindent The model Hamiltonian for the absorption algorithm is constructed as follows: first, the Hartree-Fock orbitals for the IMePh molecular system were obtained in the def2-SVP basis set and corresponding effective core potential ~\cite{peterson2003b, weigend2005a}. Next, an active space was built using the automatic valence active spaces (AVAS) method~\cite{sayfutyarova2017automated} at a threshold of $0.5$. Two chemically inspired starting points were targeted: both included the I $4d$ as well as the $5s$ and $5p$, together with O $2p$. In addition, the first space included only the C $2p_z$ orbitals, while the second included all C $2p$ orbitals. Beyond that, larger active spaces with 40 and 50 orbitals were built by adding virtual orbitals to illustrate how slowly the cost grows with increasing active space size. We emphasize that the chosen active spaces explicitly include the iodine 4$d$ semi-core orbitals, which dominate the absorption cross-section in the vicinity of 92 eV. Since EUV excitation at this photon energy primarily probes shallow-core electronic structure near the ionization threshold, inclusion of these orbitals ensures that the dominant excitation channels relevant to EUV lithography are captured within the correlated subspace. All orbital computations and active space selections were computed using the \textsc{PySCF} software library ~\cite{sun2015libcint, sun2018pyscf, sun2020recent}. Finally, within the active space chosen, compressed double factorization (CDF) of the 2-body interaction terms was introduced to reduce the overall cost of Hamiltonian simulation by this term into small fast forward-able fragments (this technique is sometimes also known as the Cartan subalgebra
approach) ~\cite{cohn2021quantum, yen2021cartan, oumarou2024accelerating}. \\ 

\subsection*{Constant-factor resource estimation for the absorption algorithm} 
The main salient parameter determining the cost of the CDF in the time evolution is the number of fragments $L$: as reasonably well-established in the literature and following Ref. \cite{fomichev2025fast}, we adopted $L = N$ as sufficient to achieve our accuracy requirements. Other important parameters for cost estimation are the effective spectral norm of the Hamiltonian $\|H\|_\omega$, the truncation and discretization of the Fourier transform integral $j_\text{max}$ and $\tau$, respectively, the desired spectral resolution determined by the broadening $\gamma$, and finally the expected magnitude of the leading Trotter error term $|\langle Y_3 \rangle |$. We will determine many of these choices in a similar way as in Ref. \cite{fomichev2025fast}. 

For the spectral norm, looking to typical sizes of EUV spectra such as in Ref. \cite{closser2017importance} of around 120 eV, or roughly 4 Ha, we adopt $\|H\|_\omega = 4$ Ha as a reasonable effective spectral range, and set $\tau = \pi / 2\|H\|_\omega$ as in Ref. \cite{fomichev2025fast}. Similarly, from that same data, a resolution of around $\gamma = 1$ eV appears acceptable for the application goal of resolving the EUV absorption sensitivity, since the spectra do not appear to vary on a much shorter energy scale than this. As shown in Ref. \cite{fomichev2025fast}, $j_\text{max} = O(\|H\|_\omega)$: since in that work it was also found that $j_\text{max} = 100$ is more than sufficient for resolving features on the order of 1 eV with $\|H\|_\omega = 2$, here we adopt $j_\text{max} = 200$. Finally, for the magnitude of the leading Trotter error $|\langle Y_3\rangle |$, we raise the already conservative estimate of 1 Ha for 18 orbitals, and increase it by another order of magnitude to 10 Ha to accommodate the larger active spaces considered here. The dependence of the Trotter time step is square-root-like due to the expression $\Delta = \sqrt{ \gamma / |\langle Y_3 \rangle | }$, so this amounts only to a modest increase to the total cost. Estimating Trotter error for these systems more firmly is an important extension that we leave to future work.

Finally, we incorporate the constant factors and shot count necessary for resolving the final absorption cross section value. As described above, the number of shots needed is
\begin{equation}
    M = (\alpha \mathcal{N}\beta/\epsilon)^2 .
\end{equation}
To estimate the computational cost of simulating the absorption sensitivity of IMePh, $\alpha = \frac{4\pi\omega}{3c}$ is photon frequency dependent prefactor, and we set the spectral bandwidth $\gamma$ to be $2\%$ of 92 eV, corresponding to experiment. The dipole normalization term, $\mathcal{N} = ||D_{\mu} \ket{\Phi}||$, depends on a variety of factors, including the physical system size, basis set, etc.  Using the Hartree-Fock ground state of IMePh over all orbitals in the def2-SVP basis, we compute the dipole normalization constant as $6.25$ a.u. for the $z$-axis, $6.94$ a.u. for the $x$-axis, and $6.84$ a.u. for the $y$-axis respectively. For the resource estimates in \Cref{tab:imeph_absorption_costs}, we use a value of $\mathcal{N} = 6.25$ for simplicity. Additionally, we target a loose screening precision of $\epsilon=0.1$, providing a relative error of approximately 10$\%$ for strong absorbers. This threshold is likely sufficient to distinguish highly EUV sensitive materials from those that are not, but distinguishing two similarly sensitive materials would require a more accurate $\epsilon$ error, and subsequently more shots. \\

\subsection*{Constant-factor resource estimation for the photoemission algorithm} 

For the results presented above, the simulation cell cubic box length has a size of $\Omega^{1/3} = 200$ Bohr, 10x the length of the longest side length of the IMePh molecule. Additionally, the state preparation cost is estimated using the \texttt{orb2mps-fq} software library~\cite{huggins2025efficient} using the IMePh molecule in an all-electron STO-3G basis set. The choice of the minimal STO-3G basis is justified by recognizing that the MPS-based orbital state preparation method is dominated by costly representations of the ``sharpness'' of core orbitals in real-space, in this case, the iodine 1$s$ orbital. The cost of the state-prep then relies solely on the number of electrons, or orbitals included and the spatial distribution of the 1$s$ orbital. 

Additionally for this calculation, the MPS optimized basis transform from Gaussians to plane-waves for each molecular orbital was ensured to have a fidelity of $\geq 0.99$, the SVD cutoff used was $10^{-3}$, and the non-Clifford gate count was computed using Eq. (F94) of Ref.~\cite{huggins2025efficient} for each MPS orbital representation, and the rotation error in the synthesis was set to $10^{-4}$. Overall, this state-preparation cost for all occupied orbitals is $\sim10^9$ non-Clifford gates, orders of magnitude lower than the Hamiltonian simulation cost in the grand total.

\section*{Data availability}

All data generated or analyzed during this study are included in this published article and its supplementary information files.

\section*{Code availability}

The resource estimation scripts used in the present study are available from the corresponding author upon reasonable request. For the pseudopotential analysis the code from \url{https://github.com/XanaduAI/pseudopotentials} was used, which is publicly available.

\section*{Competing interests}

The authors declare no competing interests.

\bibliography{Bib}

\section*{Acknowledgments}
We thank Tarik El-Khateeb for assistance with creating Fig.~\ref{fig:hero}, and Alexander Kunitsa for early feedback and helpful discussions. A part of this work was performed for Council for Science, Technology and Innovation (CSTI), Crossministerial Strategic Innovation Promotion Program (SIP), ``Promoting the application of advanced quantum technology platforms to social issues.''

\section*{Author Contributions}

T.D.K., S.F. and T.F.S. contributed to the technical work, with T.D.K. leading the development of the photoemission algorithm with significant support from T.F.S. and discussions with S.F., while S.F. focused on the absorption algorithm. T.D.K. and Q.G. together with T.F.S. and S.F. conceptualized the study with guidance from J.M.A., with T.D.K., Q.G. and S.K. providing critical input about EUV lithography and the necessary domain expertise. The manuscript was written by T.F.S., S.F. and T.D.K., and edited by all authors. All authors contributed to discussions that led to the completion of this work, and read and approved the final manuscript.

\section*{Additional information}

\textbf{Correspondence} and materials requests should be addressed to Torin F. Stetina or Stepan Fomichev.


\onecolumngrid   
\clearpage

\makeatletter
\let\SI@cite\cite
\def\cite{\let\@citex\mb@@citex \let\@newciteauxhandle\@auxoutS \SI@cite}
\def\@extra@b@citeb{@S}
\if@filesw\immediate\write\@auxoutS{\string\begingroup\string\def\string\@extra@binfo{@S}}\fi
\setcounter{NAT@ctr}{0}
\renewcommand\citenumfont[1]{S#1}
\renewcommand\@biblabel[1]{[S#1]}
\setcounter{section}{0}\renewcommand\thesection{S\arabic{section}}
\setcounter{table}{0}\renewcommand\thetable{S\arabic{table}}
\setcounter{figure}{0}\renewcommand\thefigure{S\arabic{figure}}
\setcounter{equation}{0}\renewcommand\theequation{S\arabic{equation}}
\makeatother

\begin{center}
\textbf{\large Supplementary Information for ``Quantum Simulations for Extreme
Ultraviolet Photolithography''}
\end{center}

\section{Trotter error for the absorption algorithm}
\label{app:absorption-details}

In this work, to estimate Trotter error obtaining in the implementation of the unitaries $e^{-iH\tau}$ required in the coherent time-domain algorithm, we leverage perturbative estimates and extrapolated scaling of Trotter error from Ref. \cite{fomichev2025fast}. The basic idea is based on the fact that product formulas implement exact time evolution under an approximate Hamiltonian $H' = H + \Delta ^p Y_{p+1}$. Here for the $p$th order product formula, $\Delta$ is the Trotter step and $Y_{p+1}$ is a complicated expression consisting of nested commutators of Hamiltonian CDF fragments that arises when we apply the Baker-Hausdorff-Campbell (BCH) expansion to the second-order product formula exponential decomposition. this term $Y_{p+1}$ represents the error term that arises when we employ the product formula approximation to exact time evolution. Then if we are interested in computing an  absorption spectrum, the main impact of $Y$ is to shift the peak (i.e. eigenvalue) positions relative to the true, un-perturbed Hamiltonian $H$. This peak shift may be evaluated using perturbation theory \cite{mehendale2023estimating, fomichev2025fast}: once estimated, if we impose a maximum allowable deviation $\Delta$, we can directly determine the maximum allowable Trotter step we can take -- allowing us to estimate the number of Trotter steps $r = \tau / \Delta$ we need and thus cost of implementing the time evolution unitaries needed for GQSP. 

More concretely, for the second order formula, the leading nested commutator error term is $Y_3$, given by ~\cite{fomichev2025fast}
\begin{equation}\label{eq:Y3}
 Y_3 = -\sum_{j} \Bigg[\frac{[H_j,[ \sum_{h<j}H_h, H_j]]}{24} + \frac{[\sum_{h<j} H_h,[ \sum_{h<j}H_h, H_j]]}{12} \Bigg]
\end{equation}
while the effective Hamiltonian is
\begin{equation}\label{eq:effective_Hamiltonian_U2}
        H'  = H  + \Delta^2\sum_{j} \Bigg[\frac{[H_j,[ \sum_{h<j}H_h, H_j]]}{24} + \frac{[\sum_{h<j} H_h,[ \sum_{h<j}H_h, H_j]]}{12} \Bigg] + \dots.
\end{equation}
Applying the perturbative approach to Trotter error estimation \cite{mehendale2023estimating,fomichev2025fast} lets us deduce the expression
\begin{equation}
    \label{eq:perturbation_eigenvalues}
    E'_l = E_l - \Delta^{2}\braket{E_l|Y_{3}|E_l} + O(\Delta^{4}),
\end{equation}
for eigenvalue deviation in terms of the expected magnitude of the leading Trotter error $|\langle Y_3 \rangle|$. If we choose our desired error to be $E'_l - E_l = \gamma$, which is an input to the calculation determined by the requirements of our application, we can invert this expression to determine the maximum allowable Trotter step $\Delta$, namely $\Delta = \sqrt{\gamma / |\langle Y_3 \rangle |}$. 

Combined with the description of the CDF approach, the cost of implementing the time evolution for a given time step $\tau$ can be computed in the same way as in Ref. \cite{fomichev2025fast}, with the total cost being
\begin{equation}
    \mathcal{C}_{\text{Trot}} = N_{\text{Trot calls}} \times L \times \mathcal{C}_{\text{fragment}}.
    \label{eq:perstep-cost}
\end{equation}
for $N_{\text{Trot calls}} = 2$ calls to first-order Trotter step for a second order product formula, $L$ CDF fragments, and the cost of a single fragment being given by the need to implement the $2N(N-1)/2$ Givens rotations inherent in each basis transformation, and the $2N(2N+1)/2$ $Z \otimes Z$ rotations inherent in the diagonal part of the fragment. The single-qubit rotations will be done using the phase gradient trick. The cost may be written as 
\begin{equation}
    \mathcal{C}_{\text{fragment}} = \mathcal{C}_{\text{unitary}} + \mathcal{C}_{\text{Z-matr}}, 
\end{equation}
with
\begin{align}
    \mathcal{C}_{\text{unitary}} &= 2 \frac{N(N-1)}{2} \left(2 \mathcal{C}_{\text{rot}}\right)   + 2N \mathcal{C}_{\text{rot}}, \label{eq:unitary-cost}\\
    \mathcal{C}_{\text{Z-matr}} &=  \frac{2N(2N-1)}{2} \mathcal{C}_{\text{rot}} \quad \text{or} \quad 2N \mathcal{C}_{\text{rot}}. \label{eq:zrot-cost}
\end{align}

In this study, to obtain a quick estimate of the cost, we rely on the Trotter error estimates of $|\rangle Y_3 \langle |$ made in Ref. \cite{fomichev2025fast} and take the value of $1$ Ha obtained therein for a chemical Hamiltonian with 18 spatial orbitals, and conservatively extrapolate it to 10 Ha to cover the system sizes of interest in this work. These estimates, while approximate, should be fairly robust also for the reason that that the dependence of cost on the Trotter error enters through the square-root: as such even if our estimate deviates from the true Trotter error by an order of magnitude, the expected change to the resource estimates is only a factor of three.

\section{Quantum subroutines}
\label{app:subroutines}
\subsection{Block Encoding}
\label{app-subsec:block-encode}
Block encoding is a ubiquitous protocol for implementing linear, but not necessarily unitary, operations on a quantum computer. A block encoding of a matrix $A$ can be implemented as a unitary $U_A$ by encoding the action of $A$ within a block of $U_A$. The block encoding is characterized by 3 parameters, $\alpha$ the subnormalization factor which ensures that $\norm{A}/\alpha \leq 1$, $m$ the number of ancilla qubits used to flag the successful application of $A/\alpha$, and $\epsilon$ which bounds the error between the desired operation and the block encoded matrix. The above are codified into the following definition.
\begin{defn}[$(\alpha, m, \epsilon)$-$\textsc{be}(A)$]
    Let $A$ be a matrix operating on $n$ qubits. We say that a unitary matrix $U_A$ is an $(\alpha, m, \epsilon)$-$\textsc{be}(A)$ if 
    \begin{equation}
        \norm{ (I_{2^n}\otimes \bra{0_m}) \alpha U_A (I_{2^n} \otimes \ket{0_m}) -A } \leq \epsilon,
    \end{equation}
    where $\ket{0_m}$ is the $m$ qubit all-zero state.
\end{defn}
A characterization of a block encoding at the operational level is that
\[
U_A:\ket{\psi}\ket{0_m} \rightarrow \frac{A}{\alpha}\ket{\psi}\ket{0_m} + \ket{\perp}
\]
where $\ket{\perp}$ is some unnormalized quantum state perpendicular to $\ket{\psi}\ket{0_m}$.

Throughout this work, we will employ the linear combination of unitaries (LCU) method to construct block encodings. As the name suggests, we write the block encoded matrix $A$ in terms of a linear combination of unitary matrices $U_l$, with coefficients $c_l$, i.e. $A=\sum_{l=0}^{M-1}c_lU_l$, where $m = \ceil{\log(M)}$. Without loss of generality, we may assume that the coefficients $c_l$ are positive. The subnormalization factor for the LCU approach is given by the 1-norm of the coefficients $\alpha = \sum_{l=0}^{M-1} |c_l|$.
The LCU method consists of two subroutines
\begin{equation}
    \begin{aligned}
        \textsc{prepare}:\ket{0_m} &\rightarrow \sum_{l=0}^{M-1} \sqrt{\frac{c_l}{\alpha}}\ket{l_m}\\
        \textsc{select} &= \sum_{l=0}^{M-1} U_l \otimes \ket{l_m}\bra{l_m},
    \end{aligned}
\end{equation}
consisting of a state preparation protocol on the ancilla $\textsc{prepare}$ and controlled the application of the $U_l$ in $\textsc{select}$. The block encoding $U_A$ obtained via block encoding is 
\begin{equation}
    U_A = \textsc{prepare}^\dagger \cdot \textsc{select} \cdot \textsc{prepare}.
\end{equation}

\subsection{Inequality Testing}
\label{app-subsec:ineq-test}
A common technique in this work, as well as that used in the block encoding construction of the electronic Hamiltonian, is that of \textit{inequality testing}. This technique was first introduced in Ref. \cite{sandersBlackBox}, and has been fruitfully employed to improve the efficiency of many quantum subroutines \cite{su_fault-tolerant_2021, sandersCompilationFaultTolerantQuantum2020, zini2023quantum}. The main innovation of the inequality test is the subroutine 
\begin{align*}
    \textsc{comp}:\ket{\alpha}_n\ket{\beta}_n\ket{0} \rightarrow \ket{\alpha}_n\ket{\beta}_n\ket{b(\beta < \alpha)},
\end{align*}
where $b(x)$ is a Boolean function. The implementation of the comp subroutine is simple, we just perform a subtraction circuit which requires only $n$ Toffoli gates. For our work, this subroutine naturally produces a block encoding of the position operator, since for any $\alpha \in \{0,\ldots,2^{n}-1\}$ we have
\begin{align*}
    \ket{\alpha}_n\ket{0}_n\ket{0} &\xrightarrow{\textsc{unif}} \frac{1}{\sqrt{2^n}}\sum_{\beta=0}^{2^n-1}\ket{\alpha}_n\ket{\beta}_n\ket{0}\\
    &\xrightarrow{\textsc{comp}}\frac{1}{\sqrt{2^n}}\sum_{\beta=0}^{2^n-1}\ket{\alpha}_n\ket{\beta}_n\ket{b(\beta < \alpha)}\\
    &= \frac{1}{\sqrt{2^n}}\left(\sum_{\beta \leq \alpha} \ket{\alpha}_n\ket{\beta}_n\ket{0} + \sum_{\beta>\alpha}\ket{\alpha}_n\ket{\beta}_n\ket{1}\right)\\
    &\xrightarrow{\textsc{unif}^\dagger} \frac{1}{2^n}\ket{\alpha}_n\ket{0}_n\left(\ket{0}\sum_{0 \leq x \leq \alpha -1 } + \ket{1}\sum_{\alpha - 1\leq x \leq 2^n -1}\right)\\
    &=\frac{\alpha}{2^n}\ket{\alpha}_n\ket{0}_n\ket{0} + \frac{2^n - \alpha}{2^n}\ket{\alpha}_n\ket{0}_n\ket{1}.
\end{align*}
Since the quantum states are encoded with signed integers, we would just need to apply a Pauli Z controlled on the sign bit to get the signed position operator we use. 

The inequality testing protocol can be extended to compute more complex functions as well. For the projector onto the continuum states, we need to test if $\norm{q_i} \leq R_c$. Since $R_c$ is a fixed constant, we do not need to load it into memory. Since the values stored in binary are the indexes to the positions in real space, we need to scale the radius by the lattice spacing $h = \frac{N^{1/3}}{\Omega^{1/3}}$, and compute a binary representation of  $\left(\frac{R_c}{h}\right)^2$. Now our goal is to determine if the inequality
\[
\norm{q_l}^2 \leq \left(\frac{R_c}{h}\right)^2
\]
on the quantum computer. We can do this by first computing $\norm{q_l}^2$ in binary, and replacing $\textsc{comp}$ with a subtraction by classical constant circuit, which we will denote as $\textsc{comp}_c$. The operations on the quantum computer are as follows
\begin{align*}
    \ket{q_x}\ket{q_y}\ket{q_z}\ket{0}\ket{0} &\xrightarrow{\textsc{sum-squares}}\ket{q_x}\ket{q_y}\ket{q_z}\ket{\norm{q}^2}\ket{0}\\
    &\xrightarrow{\textsc{comp}_c} \ket{q_x}\ket{q_y}\ket{q_z}\ket{\norm{q}^2}\ket{b(\norm{q}^2 \leq R^2},
\end{align*}
and uncomputing the subtraction by classical constant and sum-of-squares circuits, we obtain the state 
\[
\ket{q_x}\ket{q_y}\ket{q_z}\ket{0}\ket{b(\norm{q}^2 \leq R^2},
\]
which provides a single qubit flag denoting the status of the inequality test. 

This technique is particularly powerful when combined with the construction of Ref. \cite{gidneyHalvingCostQuantum2018}. This paper introduces a technique called ``measure and fixup'', which replaces Toffolis used in the uncomputation of a circuit with a measurement and a controlled single qubit Pauli $Z$ operation. Thus, the circuits for squaring, \textsc{comp}, and other arithmetic circuits can be uncomputed with zero non-Clifford cost, essentially halving the overall cost of the subroutine.

\section{Block Encoding of the First quantized Plane Wave Hamiltonian}
\label{app:BE-H}
The block encoding of the first quantized plane wave Hamiltonian was described in great detail in Ref. \cite{su_fault-tolerant_2021}. We review the main subroutines and costs associated with constructing this block encoding. First, we provide a restatement of Theorem 4 of Ref. \cite{su_fault-tolerant_2021}, which gives the overall Toffoli gate cost to perform the block encoding.

\begin{theorem}[Resource estimate to block encode $H$]
\label{thm:rsrc-est-be-H}
    Let $H$ be the electronic Hamiltonian describing a system of $\eta$ electrons occupying positions in a domain $\Omega \subset \mathbb{R}^3$, discretized using $N^{1/3} \in \mathbb{N}$ plane waves per degree of freedom. Let $n = \ceil{\log(N^{1/3})}$ be the number of qubits per degree of freedom, for a total of $3\eta n$ qubits to describe all of the system degrees of freedom.
    Then, using the methods described in Ref. \cite{su_fault-tolerant_2021}, one can obtain an $(\lambda, n_\text{ref},\epsilon)$-$\be(H)$ where 
    \begin{equation}
    \begin{aligned}
        \lambda &=  \max\{\lambda_T' + \lambda_U +\lambda_V, \frac{\lambda_U + \lambda_V(1-\eta^{-1})^{-1}}{p_\nu}\}\\
        n_\text{ref} &= 3n^2 + 4n_M(n+1) + 6n + 5 + \max\{n_T, n_R+1\},
    \end{aligned}
    \end{equation}
    and $p_\nu$ is the probability of preparing the momentum state $\sum_{\nu \in G_0}\frac{1}{\norm{\nu}}\ket{\nu}$, and $\lambda_T' = \frac{6\eta \pi^2}{\Omega^{2/3}}2^{2(n-1)}$.
    With $n_T$, $n_R$, $n_M$ chosen according to
    \begin{equation}
    \begin{aligned}
        \epsilon_M &= \frac{2\eta}{2^{n_M}\pi \Omega^{1/3}}(\eta - 1 + 2\lambda_\zeta)\\
        \epsilon_R &= \frac{\eta \lambda_\zeta}{2^{n_R}\Omega^{1/3}}\sum_{\nu \in G_0}\frac{1}{\norm{\nu}}\\
        \epsilon_T &= \frac{\pi \lambda}{2^{n_T}},
    \end{aligned}
    \end{equation}
    so that overall
    \begin{equation}
        (\epsilon_M + \epsilon_R + \epsilon_T)^2 < \epsilon^2.
    \end{equation}

    Moreover, this block encoding is accomplished using 
    \begin{equation}
        n_M + 6n + 2n_{\eta} + n_{\eta \zeta} + \max\{5n_R -4, 5n +1\} +18
    \end{equation}
    ancilla qubits in addition to $n_r$. The Toffoli cost to perform one query to the qubitized block encoding is 
    \begin{equation}
    \begin{aligned}
    \label{eq:toff-comp-BE}
        2&(n_T + 4n_{\eta \zeta} + 2b_r - 12) + 14n_\eta + 8b_r - 36 + 12\eta n + 4\eta -8 + 5(n-1) +2 + \\
        &3n^2 + 15n + 4 n_M(n+1) - 7 + \lambda_\zeta + 24n + 3(2n n_R - n(n+1) -1).
    \end{aligned}
    \end{equation}
\end{theorem}

\begin{table*}[ht]
    \centering
\begin{tblr}{
  colspec = {|c|Q[c,m]|Q[c,wd=6cm]|},
  rowspec = {|Q[m]|Q[m]|Q[m]|},
  vlines,
  hlines,
  }
\SetCell[c=3]{c} \textbf{Quantities characterizing Hamiltonian block encoding costs} \\
Symbol & Scaling & Description \\
$N$ & $\Omega k_{\max}^{3/2}$ & Number of plane wave modes per electron\\
$\Omega$ & Problem dependent ($200^3 \text{\AA}^3$ here)   & The volume of the computational region\\
$n_M$ & $\log\left(\frac{4\eta^2}{\Omega^{1/3}}\right)$ & Number of ancilla qubits in inequality testing for preparation of Coulomb potential \\
$n_R$ & $\log\left(\lambda_U\right)$ & Number of ancilla qubits to encode nuclear coordinates in binary \\
$n_T$ & $\log\left(\lambda\right)$ & The number of ancilla qubits used to encode the coefficient selecting between $T$ and $U+V$ \\
$n_\eta$ & $\log(\eta)$ & Number of ancilla qubits to encode $\eta$ degrees of freedom\\
$n_{\eta \zeta}$ & $\log(\eta + 2\lambda_\zeta)$ & Number of ancilla qubits to encode $\eta + 2 \lambda_\zeta$ degrees of freedom\\
$b_r$ & $O(1)$ & Number of bits of precision in rotation angle for preparing equal superposition state\\
$\lambda_\zeta$ & $\sum_{j=0}^{L-1} |\zeta_j|$  & One norm of nuclear charges\\
$\lambda_T$ & $6\eta\pi^2 \frac{N^{2/3}}{\Omega^{2/3}}$  & Subnormalization factor for the kinetic energy term\\
$\lambda_U$ & $\eta\lambda_\zeta\frac{N^{1/3}}{\Omega^{1/3}}$ & Subnormalization factor for nuclear-electron interaction \\
$\lambda_V$ & $\eta^2\frac{N^{1/3}}{\Omega^{1/3}}$ & Subnormalization factor for electron-electron interaction 
\end{tblr}
\caption{The quantities characterizing the costs in the block encoding of the electronic Hamiltonian, under Born-Oppenheimer approximation, in a first quantized plane wave basis.}
\label{tab:quantities-in-BE-H}
\end{table*}

\begin{cor}[Block encoding of IMePh Hamiltonian]
\label{cor:BE-imeph-naiv}
    The IMePh molecule has $\eta = 110$ electrons and $\lambda_\zeta = 110$. Choosing a side length $\Omega^{1/3}$ to be $\sim 10\times$ the length of longest side of the molecule side gives $\Omega^{1/3} = 200$ \AA. The kinetic energy of a grid of $N^{1/3}$ points per dimension can represent a kinetic energy $E_{cut} = \frac{1}{2}\frac{N^{2/3}}{\Omega^{2/3}}$. Since the $1$s orbital of Iodine has kinetic energy of $\sim 1215$Ha, we can safely assume that the largest kinetic energy of the molecule $T_{max} < \eta \times 1215$Ha. Therefore, we may choose a number of grid points 
    \begin{equation}
        N^{1/3} \geq \frac{200}{\pi}\sqrt{\frac{2 \times 110\times 1215}{3}} \sim 1.9 \times 10^4,
    \end{equation}
    and 
    \begin{equation}
        n = \ceil{\log(N^{1/3})} = 15
    \end{equation}
    qubits to represent each degree of freedom. Therefore, we require $3 n \eta = 4950$ qubits to represent the system. The one norms of the relative terms are 
    \begin{equation}
    \begin{aligned}
        \lambda_T' &= 4.37 \times 10^7\\
        \lambda_U &= 7.5 \times 10^6\\
        \lambda_V &= 1.23 \times 10^7
    \end{aligned}
    \end{equation}
    and find that $P_{eq} \sim 1$, and the overall subnormalization factor is 
    \begin{equation}
        \lambda \sim  5.00\times 10^7.
    \end{equation}
    
    Taking the overall $\epsilon = .5\epsilon_{chem} \sim 8 \times 10^{-4} \text{Ha}$, and taking the choice of $\epsilon_T = \epsilon_R = \epsilon_M = \epsilon/3$, gives 
    \begin{equation}
    \begin{aligned}
        n_M &= 36\\
        n_R &= 50\\
        n_T &= 35.
    \end{aligned}
    \end{equation}
    Therefore, the total number of Toffoli gates to perform one query of the block encoding is 
    \begin{equation}
        T_1 =  2.53 \times 10^4,
    \end{equation}
    and the number of ancilla qubits is 
    \begin{equation}
        q_{anc} = 3542,
    \end{equation}
    for a total of 
    \begin{equation}
        q_{tot} = 8084
    \end{equation}
    logical qubits in the simulation.
\end{cor}

\section{Pseudopotential modified FQ method}\label{app:pseudo}

While pseudopotential usage is rapidly becoming routine in quantum algorithms, it is still a highly demanding task to re-build a given algorithm with pseudopotentials in mind. In this manuscript, we aim to only obtain a rough estimate of the cost reduction to be expected from the use of pseudopotentials as opposed to all-electron calculations, and reserve the full derivation of the implementation of the pseudopotential for this case to a follow-up work.

The basic motivation to use pseudopotentials is that EUV light with ~92 eV energy will not excite deep-core electrons, so they might as well be frozen out in the pseudopotential. However, care needs to be taken with this, because EUV light will excite shallow-core (or semi-core) electrons, unlike what happens in, for example, UV-Vis spectroscopy. For the IMePh molecule with iodine, formal valence counting suggests that the iodine valence orbitals are the $5s$ and $5p$, leaving only 7 valence electrons outside the core: however, it is well-known that the semi-core $4d$ orbitals play a key role in EUV absorption \cite{closser2017importance}. This can be gleaned from the fact that the $4d$ (atomic-like) binding energies are around 50 eV based on tabulated X-ray excitation data \cite{booklet2001x}, so they can easily be excited by EUV at 92 eV. By contrast, the next-deepest levels of $4s$/$4p$ have binding energies in excess of 120 eV, so these and deeper levels can likely be frozen out.

The challenge is that most standard pseudopotentials for iodine \cite{hartwigsen1998relativistic} freeze out $4d$ orbitals, and are thus not directly suitable. There do exist plane-wave based, norm-conserving pseudopotentials that do keep those $4d$ electrons (and also optionally the $4s$ and $4p$) explicitly \cite{van2018pseudodojo}. At the same time, there is an indication that electron dynamics and spectra can be calculated with high accuracy using such semi-core pseudopotentials, as evidenced by a recent benchmark of computing L-edge XANES for Al$_2$O$_3$ with $2s$ and $2p$ aluminum semi-core orbitals using TDDFT \cite{urquiza2023pseudopotential}. While this is not precisely the calculation we intend to run, this is a strong signal that carefully chosen pseudopotentials that include semi-core orbitals allow accurate calculation of dynamical properties like spectra, while freezing out deep core electrons and thus allowing to reduce the algorithmic cost.

The analysis above suggests that $4d$ semicore pseudopotentials for iodine exist. Unfortunately, that particular type of pseudopotential is not analytic, but fully numerical -- the values of the pseudopotential and pseudowavefunctions are tabulated at each spatial point rather than expressed in terms of an analytical form. Since all the existing methods for implementing pseudopotentials in quantum algorithms rely on having an analytic form, this presents a further challenge to our resource reduction estimation process. 

We approach this in the following way: we combine the widespread existence of analytic pseudopotentials \cite{hartwigsen1998relativistic}, together with the evidence above about the existence of a high-quality numerical pseudopotential with $4d$ semicore inclusion \cite{van2018pseudodojo}, to assume that an analytic-style pseudopotential of good-enough quality with suitable inclusion of $4d$ may be constructed. As a stand-in for such a pseudopotential, we use the coefficients obtained in Ref. \cite{hartwigsen1998relativistic} for the key elements in the IMePh molecule: for the case of iodine, we consider the coefficients as given in the table in the reference, assuming that the only changes expected from the semicore $4d$ are the additional valence electrons -- for a total of 17 -- and perhaps some minor modification of the projector coefficients, to which the resource estimation is only weakly sensitive. 

With this assumption, we turn to the pseudopotential implementation method described in Ref. \cite{zini2023quantum} and specifically to the accompanying public Github repository \url{https://github.com/XanaduAI/pseudopotentials} for estimating the resource reduction. 

We need one final assumption to make the estimates. The method of Ref. \cite{zini2023quantum} only implements a single projector per angular momentum direction. For the IMePh system, based on the coefficients given in Ref. \cite{hartwigsen1998relativistic}, all the elements include only a single projector per angular momentum except iodine: we assume that the cost of the block-encoding will not be significantly affected by this, which is expected from the literature. 

With these assumptions in place, we can directly leverage the parameters of the IMePh system studied here -- the unit cell size, chemical content, number of valence electrons of each constituent element, the projector coefficients -- to make the estimates using the code in \url{https://github.com/XanaduAI/pseudopotentials}. These estimates are presented in the main text.  

\section{IMePh geometry}

Below is the table providing the XYZ Cartesian coordinates for the IMePh geometry used in the present study.

\begin{table}[htbp]
  \centering
  \caption{Optimized geometry (\texttt{def2-SVP} with \texttt{def2-SVP} ECP on I,
           $\omega$B97X-D functional, geomeTRIC/PySCF).}
  \label{tab:xyz-geometry}
  \begin{tabular}{c r r r}
    \hline
    Atom & $x$ (\AA) & $y$ (\AA) & $z$ (\AA) \\
    \hline
    C & $-0.13355$ & $ 0.10021$ & $-0.00000$ \\
    C & $ 1.27216$ & $ 0.04122$ & $-0.00000$ \\
    C & $ 1.98181$ & $ 1.24148$ & $-0.00000$ \\
    C & $ 1.32201$ & $ 2.47281$ & $-0.00000$ \\
    C & $-0.06787$ & $ 2.52180$ & $-0.00000$ \\
    C & $-0.79156$ & $ 1.33025$ & $-0.00000$ \\
    O & $-0.79728$ & $-1.07819$ & $-0.00000$ \\
    H & $-1.74544$ & $-0.91788$ & $-0.00000$ \\
    H & $ 3.07358$ & $ 1.20738$ & $-0.00000$ \\
    H & $-0.59536$ & $ 3.47698$ & $-0.00000$ \\
    H & $-1.88530$ & $ 1.36177$ & $-0.00000$ \\
    C & $ 1.96010$ & $-1.29500$ & $-0.00000$ \\
    H & $ 1.67251$ & $-1.88633$ & $-0.88249$ \\
    H & $ 1.67251$ & $-1.88633$ & $ 0.88249$ \\
    H & $ 3.05182$ & $-1.17853$ & $-0.00000$ \\
    I & $ 2.43809$ & $ 4.25776$ & $-0.00000$ \\
    \hline
  \end{tabular}
\end{table}

\makeatletter
\newwrite\SI@notes
\immediate\openout\SI@notes=SNotes.bib
{\let\@bibdataout\SI@notes \let\@auxout\@auxoutS \@bibdataout@rev}
\immediate\closeout\SI@notes
\begingroup
 \let\SI@label\label \def\label#1{\SI@label{S#1}}
 \def\@bibsetup#1{\setlength{\topsep}{\z@}\NATx@bibsetnum{\ref{SLastBibItem}}}
 \bibliographystyleS{apsrev4-2}
 \bibliographyS{Bib}
\endgroup
\if@filesw\immediate\write\@auxoutS{\string\endgroup}\fi
\makeatother

\end{document}